\documentclass[preprint,
                preprintnumbers,
                amsmath,
                amssymb,
                superscriptaddress,
                a4paper]{revtex4}

\usepackage{latexsym}
\usepackage{amsfonts}
\usepackage{amsmath}
\usepackage{bm}
\usepackage[dvips]{graphicx}
\usepackage{color}
\usepackage[utf8]{inputenc}
\newcommand{\placefig}[1]{
    \begin{center}
     {\bf FIGURE \ref{#1} AROUND HERE}
    \end{center}
}

\newcommand{\placetab}[1]{
    \begin{center}
     {\bf TABLE \ref{#1} AROUND HERE}
    \end{center}
}

\DeclareMathAlphabet{\mathpzc}{OT1}{pzc}{m}{it}

\def\figfoot{Vendrell and Meyer, Journal of Chemical Physics}

\newcommand{\figcaption}[2]{
    \noindent {\bf Figure \ref{#1}:} #2
    \vspace{1cm}
}

\newcommand{\z}{\ensuremath{{l;\kappa_1,\ldots,\kappa_l}}}
\newcommand{\zll}{\ensuremath{{l+1;\kappa_1,\ldots,\kappa_{l+1}}}}

\newcommand{\zn}{\ensuremath{{l;\kappa_1,\ldots,\kappa_{l-1}}}}

\newcommand{\znm}{\ensuremath{{l-1;\kappa_1,\ldots,\kappa_{l-2}}}}
\newcommand{\Zl}{\ensuremath{{z,\kappa_l}}}
\newcommand{\Zll}{\ensuremath{{z\!+\!1,\kappa_{l+1}}}}
\newcommand{\Zlll}{\ensuremath{{z\!+\!2,\kappa_{l+2}}}}
\newcommand{\Zml}{\ensuremath{{z\!-\!1,\kappa_{l-1}}}}

\begin{document}

\author{Oriol Vendrell}
\email[e-mail: ]{oriol.vendrell@cfel.de}
\affiliation{Center for Free-Electron Laser Science, DESY,
             Notkestrasse 85, D-22607 Hamburg, Germany}

\author{Hans-Dieter Meyer}
\email[e-mail: ]{hans-dieter.meyer@pci.uni-heidelberg.de}
\affiliation{Theoretische Chemie, Physikalisch-Chemisches Institut,
   Universit\"at Heidelberg, INF 229,
   D-69120 Heidelberg, Germany}

\title{Multilayer multi-configuration time-dependent Hartree method: 
implementation and applications to a Henon-Heiles Hamiltonian and to
pyrazine.}

\date{\today}

\begin{abstract}
  The multilayer multiconfiguration time-dependent Hartree (ML-MCTDH) method is
  discussed and a fully general implementation for any number of layers based
  on the recursive ML-MCTDH algorithm given by Manthe [J. Chem. Phys. {\bf
  128}, 164116 (2008)] is presented.  The method is applied first to a
  generalized Henon-Heiles (HH) Hamiltonian.  For 6D HH the overhead of
  ML-MCTDH makes the method slower than MCTDH, but for 18D HH ML-MCTDH starts
  to be competitive.  We report as well 1458D simulations of the HH Hamiltonian
  using a seven layer scheme. The photoabsorption spectrum of pyrazine computed
  with the 24D Hamiltonian of Raab {\em et. al.} [J. Chem. Phys. {\bf 110}, 936
  (1999)] provides a realistic molecular test case for the method. Quick and
  small ML-MCTDH calculations needing a fraction of the time and resources of
  reference MCTDH calculations provide already spectra with all the correct
  features. Accepting slightly larger deviations, the calculation can be
  accelerated to take only 7 minutes.  When pushing the method towards
  convergence, results of similar quality than the best available MCTDH
  benchmark, which is based on a wavepacket with  $4.6\times 10^7$
  time-dependent coefficients, are obtained with a much more compact
  wavefunction consisting of only $4.5\times 10^5$ coefficients and requiring a
  shorter computation time.
\end{abstract}

% --- Begin article ---

\maketitle

\section{Introduction} \label{sec:introduction}

  % General bla how important is quantum dynamics and MCTDH
  The multiconfiguration time-dependent Hartree (MCTDH) method
 \cite{mey90:73,man92:3199,bec00:1,mey03:251,mey09:book} has become over
  the last decade the tool of choice to accurately describe the dynamics of
  complex multidimensional quantum mechanical systems, providing in many cases
  reference results that are used to benchmark other approaches or determine
  the behavior of approximate methods.
  MCTDH was initially formulated with molecular quantum dynamics in mind, where
  the degrees of freedom are distinguishable and the potential operator
  correlates in principle all vibrational degrees of freedom of the system.
  Many successful applications over the years deal with the spectroscopy of
  molecules~\cite{raa99:936,mar05:204310,dor08:224109,wor08:569,ven09:034308},
  isomerisation and intramolecular vibrational energy redistribution
  (IVR)~\cite{ric04:6072,iun04:6992}, inelastic and reactive scattering
  calculations~\cite{suk02:10641,ott08:064305,wu04:2227,har07:084303,%
  bha10:214304}, scattering of atoms or molecules at
  surfaces~\cite{hei01:1382,har04:3829,cre06:074706}, etc.
  More recently, the potential of the MCTDH method for the treatment of systems
  of indistinguishable particles, either fermions or bosons, has been realized
  as well, and many applications can be counted nowadays in such new
  directions~\cite{sak09:2206011,alo09:301,zoe08:040401,zan04:763,%
  jor08:025035,nes07:214106,suk09:223002}.
  Although the equations of motion remain the same in such cases,
  the symmetry properties of the wavefunction and the often two-body nature of
  the interactions has lead to the appearance of dedicated programs that
  specifically and efficiently deal with such
  cases~\cite{zan03:1064,nes05:124102,alo08:033613}.

  In general, the solution of the time-dependent Schr{\"o}dinger equation in
  a direct-product basis of 1D functions (primitive functions) scales
  exponentially as $N^f$, where $N$ is the number of primitive basis functions
  per degree of freedom and $f$ is the number of degrees of freedom. In the
  MCTDH {\em ansatz} one introduces an optimal time-dependent (TD) basis
  (called SPF for single particle function) for each degree of freedom, which
  can be kept smaller than the underlying primitive basis, leading to a better
  scaling of the number of configurations $n^f$.  In this way, MCTDH can deal
  with larger systems than the standard method, although the exponential
  scaling with $f$ remains, but to a lower base. The base can be further
  reduced by grouping degrees of
  freedom together in what are called combined modes or logical coordinates.
  One obtains in this way a smaller number of effective degrees of freedom $p$,
  leading to a smaller number of configurations, but the TD basis
  functions that have to be now propagated are multidimensional. For a given
  problem, it is possible to find mode-combination schemes that provide an
  optimum balance between the effort of propagating the expansion coefficients
  of each configuration (the A-vector in usual MCTDH terminology) and the
  effort of propagating the multidimensional SPFs . With
  increasing dimensionality, however, the multidimensional SPFs,
  the A-vector, or both, become harder and harder to propagate. As a rule of
  thumb and very generally, the number of degrees of freedom that can be
  treated for a correlated vibrational problem nowadays is about 20, but for
  systems well suited for MCTDH more than 60 degrees of freedom (DOF) have been
  accounted for~\cite{wan00:9948,nes03:24}.

  % Introduce multilayer
  Multilayer (ML) MCTDH represents a powerful extension of the usual MCTDH
  approach, in which a multiconfigurational {\em ansatz} is used for the
  multidimensional SPFs themselves.  This results in an extra layer of TD
  coefficients that have to be propagated and hence the name.  One way to think
  of the ML wavefunction {\em ansatz} (given below) is to see the standard
  propagation method and MCTDH as particular cases of a ML wavefunction. In the
  standard propagation method a single layer of expansion coefficients with
  respect to some time-independent (TI) basis is present. MCTDH contains a first
  layer of expansion coefficients with respect to the SPF basis, and a second
  layer of TD expansion coefficients that parameterize the time evolution of the
  SPFs. Expanding multidimensional SPFs using a MCTDH-like {\em ansatz} one
  obtains a three-layer scheme.  Schemes with more than three layers are of
  course possible. In passing we note that the ML-MCTDH {\em ansatz} can
  be connected to tensor contraction techniques in the mathematical
  literature~\cite{chi07:084110} and to density matrix renormalization
  group (DMRG) theory~\cite{whi99:4127}.

  ML-MCTDH was first formulated by Wang and Thoss, who also provided an
  implementation of the method for three layers (in the present paper MCTDH
  is regarded as a two-layer scheme) and showed its applicability on the
  spin-boson model and on an electron-transfer model~\cite{wan03:1289}. A
  formally identical formulation, but without an implementation, was given
  independently at the same time by Meyer and Worth~\cite{mey03:251},
  who termed it cascading MCTDH.
  Over the last years Wang and Thoss have further developed ML-MCTDH, expanding
  the number of layers that their code can handle and treating condensed phase
  model Hamiltonians of large dimensionality between 100 and a few thousand
  DOF~\cite{wan06:034114,cra07:144503,kon07:11970,wan08:115005}. The same authors also described how
  ML-MCTDH can be used to deal with identical particles, which is accomplished
  defining the problem in Fock space and directly working in the
  occupation-number representation~\cite{wan09:024114}. (Note that ML-MCTDH
  cannot be applied to identical particles in the usual first-quantized
  formulation because the grouping of DOF into combined modes destroys the
  necessary exchange symmetry properties of the wavefunction.)
  Recently, Manthe provided a recursive formulation of the ML-MCTDH equations
  of motion (EOM) for any number of layers and described the recursive
  algorithm that has to be used to compute all quantities entering the
  EOM~\cite{man08:164116,man09:054109}.

  % Motivate this paper
  In this contribution, we present the implementation of the EOM and recursive
  algorithm proposed by Manthe, and discuss its applicability and performance
  on a Henon-Heiles (HH) model system and on a realistic non-adiabatic molecular
  case, full dimensional 24D pyrazine. The resulting implementation is fully
  general and it can handle standard method (one layer), normal MCTDH (two
  layers) and any desired multilayering scheme. In this paper, we report on
  simulations with three layers and up to a seven-layer case for the 1458D
  HH Hamiltonian. We also simulate pyrazine using five and six layer
  schemes.
  The ML-MCTDH concept, EOM and recursive algorithm are discussed first,
  emphasizing some aspects of our implementation such as the treatment of the
  separable parts of the Hamiltonian. We have tried to keep such discussion
  self-contained and to reflect the way our code operates, but for an in-deep
  discussion Ref.~\cite{man08:164116} should be read. The performance and
  correctness of the recursive implementation are tested on the HH
  Hamiltonian. For this system, 6D calculations with varying coupling strength
  are carried on first, which are directly compared to MCTDH results. For such
  a low dimensionality, the algorithmic complexity of ML-MCTDH does not yet pay
  off, and MCTDH is more efficient than ML-MCTDH.  For 18D HH we find
  that ML-MCTDH starts to be competitive. For this case we analyze the
  convergence properties of the method with respect to the number of basis
  functions at each layer.
  Simulations of 1458D HH are also reported, showing the
  power of this approach. The ML-MCTDH implementation is then applied to the
  photoabsorption spectrum of pyrazine using the 24D vibronic-coupling
  Hamiltonian of Raab and coworkers~\cite{raa99:936}. This is a realistic
  molecular Hamiltonian involving the presence of a conical intersection
  between the electronic state $S_2$ initially populated by photoabsorption and
  the $S_1$ electronic state to which the system decays within some ten fs.
  The Hamiltonian of Ref.~\cite{raa99:936} has been used to test several
  quantum dynamical approaches. In the case of ML-MCTDH, we find that the
  method performs very efficiently and provides already well converged spectra in
  about a fifth of the CPU time used by the smallest of the MCTDH reference
  results using extremely few computational resources. Tolerating a larger
  error one can reduce the computation time to 7 minutes and still get a
  spectrum with all basic features in place. This is amazingly fast,
  considering that one is dealing with a fully correlated 24-dimensional
  quantum dynamics calculation. No other quantum dynamical method applied up to
  date to the pyrazine Hamiltonian offers the extreme quality/cost relation
  provided by ML-MCTDH. When pushing ML-MCTDH towards convergence with the
  pyrazine Hamiltonian, we find that it still offers results of at least the
  same quality (if not better) than the best benchmark results available from
  MCTDH calculations with about $4\times 10^7$ configurations, at a lower
  computational cost.

  % Describe the contents of this paper
  The paper is organized as follows: Section \ref{sec:theory} discusses the
  ML-MCTDH equations and implementation. Section \ref{sec:hh} discusses the
  results on the HH Hamiltonian and Section \ref{sec:pyr} discusses
  the application of ML-MCTDH to the photoabsorption spectrum of pyrazine.
  Section \ref{sec:conclusion} summarizes the results and provides some future
  perspectives for the method.

\section{Multilayer MCTDH} \label{sec:theory}

  \subsection{{\em Ansatz} and general concept}
  % ML ansatz
  The wavefunction of a system of distinguishable degrees of freedom, e.g.
  representing molecular vibrations, can be conveniently represented as an
  expansion in terms of direct products of orthonormal basis functions, one for
  each degree of freedom
  \begin{equation} \label{eq:ansatz1}
    \Psi(q_1,\ldots,q_f,t) = \sum_{j_1}^{N_1} \cdots \sum_{j_f}^{N_f}
      A_{j_1,\ldots,j_f}^1(t)
      \cdot \chi_{j_1}^{(1)}(q_1)\cdot \ldots \cdot \chi_{j_f}^{(f)}(q_f).
  \end{equation}
  Following the notation introduced by Manthe~\cite{man08:164116}, the
  $A_{j_1,\ldots,j_f}^1$ denote the TD coefficients of the
  first (and by now the only) layer of TD expansion terms.

  In the MCTDH approach, the wavefunction is expanded in terms of direct
  products of orthonormal {\em time-dependent} single particle functions (SPF)
  \begin{equation} \label{eq:ansatz2}
    \Psi(q_1,\ldots,q_f,t) = \sum_{j_1}^{n_1} \cdots \sum_{j_f}^{n_f}
      A_{j_1,\ldots,j_f}^1(t)
      \cdot \varphi_{j_1}^{(1;1)}(q_1,t)\cdot \ldots \cdot \varphi_{j_f}^{(1;f)}(q_f,t),
  \end{equation}
  which are themselves expanded in terms of the underlying primitive basis
  \begin{equation} \label{eq:ansatz3}
    \varphi_{m}^{1;\kappa}(q_\kappa,t) = \sum_{j}^{N_\kappa}
      A_{m;j}^{2;\kappa}(t) \cdot \chi_{j}^{(\kappa)}(q_\kappa).
  \end{equation}
  Therefore, MCTDH can be seen as a 2-layer scheme with TD coefficients
  $A_{j_1,\ldots,j_f}^1(t)$ at the top layer, and sets of second layer TD
  coefficients $A_{m;j}^{2;\kappa}(t)$ for each degree of freedom. We usually
  refer to the 1-layer scheme as the standard method, to the 2-layer scheme simply
  as MCTDH, and to deeper layering schemes as ML-MCTDH.
  Note the important detail that all SPFs of a certain degree of freedom are
  expanded in terms of the same underlying basis, i.e. the
  $\chi_{j}^{(\kappa)}(q_\kappa)$ functions in Eq.~(\ref{eq:ansatz3}) have no
  $m$ index.

  The computational gain of MCTDH with respect to the standard method arises
  from the expansion orders $n_\kappa$ being in general smaller than the size of
  the underlying primitive basis $N_\kappa$, which leads to a smaller number of
  TD coefficients to be propagated. However, the total number of TD
  coefficients is still given by $\prod_{\kappa=1}^f n_\kappa$, and therefore
  the computational effort raises exponentially with the number of degrees of
  freedom. Hence, MCTDH does not eliminate the exponential scaling but reduces
  the base to which the scaling occurs (for a detailed analysis of the
  computational scaling of MCTDH see for example Ref.~\cite{bec00:1}).
  The base can be further reduced by combining the physical coordinates
  $q_1,\ldots,q_f$ into logical coordinates (also referred to as combined
  modes) $Q_1^1,\ldots,Q_p^1$, such that each logical coordinate comprises one
  or several of the physical coordinates, as $Q_\kappa^1
  =\{q_{1_\kappa},\ldots,q_{d_\kappa}\}$. Here a more general notation to
  designate a combined coordinate has been introduced, that will be useful when
  discussing deeper layering schemes.  Mode combination has been extensively
  used in many applications of the MCTDH method. Similarly as in
  Eq.~(\ref{eq:ansatz2}), the MCTDH wavefunction now reads
  \begin{equation} \label{eq:ansatz4}
    \Psi(Q_1^1,\ldots,Q_p^1,t) = \sum_{j_1}^{n_1} \cdots \sum_{j_p}^{n_p}
      A_{1;j_1,\ldots,j_p}^1(t)
      \cdot \varphi_{j_1}^{(1;1)}(Q_1^1,t)\cdot \ldots \cdot \varphi_{j_p}^{(1;p)}(Q_p^1,t),
  \end{equation}
  and the TD basis functions $\varphi_{j_\kappa}^{(1;\kappa)}(Q_\kappa^1,t)$ are now
  multidimensional. Introducing mode-combination,
  the computational effort is switched from the propagation of a large
  vector of $A_{j_1,\ldots,j_p}^1(t)$ coefficients and one-dimensional SPFs, to
  a shorter vector of coefficients but multi-dimensional, harder to propagate
  SPFs. It is often the case, that some experience and knowledge of the system
  under consideration is necessary to find an efficient mode-combination scheme
  for a given problem.
  The mode-combined SPFs are given by
  \begin{equation} \label{eq:ansatz5}
    \varphi_{m}^{1;\kappa}(Q_{\kappa}^1,t) =
    \sum_{j_{1_\kappa}}^{N_{1_\kappa}} \cdots \sum_{j_{d_\kappa}}^{N_{d_\kappa}}
     A_{m;j_{1_\kappa},\ldots,j_{d_\kappa}}^{2;\kappa}(t)\cdot
     \chi_{j_{1_\kappa}}^{(\kappa,1)}(Q_{1_\kappa}^{2;\kappa}) \cdot\ldots\cdot
     \chi_{j_{d_\kappa}}^{(\kappa,d_\kappa)}(Q_{d_\kappa}^{2;\kappa}),
  \end{equation}
  where the last equation resembles the {\em ansatz} in Eq.~(\ref{eq:ansatz1})
  in that multidimensional SPFs are represented as a multiconfigurational
  expansion in terms of underlying time-independent basis functions.
  In the same way a TD basis was introduced above in going from the standard
  {\em ansatz} to MCTDH, effectively adding a second layer of TD expansion
  coefficients, an MCTDH expansion can be used to represent the
  $\varphi_{m}^{1;\kappa}(Q_\kappa^1,t)$ SPFs, effectively adding a third layer of
  expansion coefficients. This results in a 3-layer ML-MCTDH {\em ansatz}.

  The ML-MCTDH layering scheme can be very flexible. In a high dimensional
  system one combines groups of degrees of freedom into high dimensional SPFs
  until the size of the vector of coefficients in Eq.~(\ref{eq:ansatz4})
  becomes manageable. However, the combined SPFs are too large to be
  efficiently propagated. Then, one breaks the combined modes into even smaller
  groups of logical coordinates introducing a new layer of coefficients, whose
  size is manageable. This procedure can be repeated over and over until the
  (possibly combined) primitive degrees of freedom are reached.

  Owing to the flexibility of the layering scheme, and the fact that ML-MCTDH
  wavefunctions can be many layers deep, it is convenient to introduce a
  diagrammatic notation to represent these objects~\cite{man08:164116}.  In
  this notation, wavefunctions are represented by {\em trees}, i.e. connected
  graphs without loops. Each node in the tree represents a set of vectors of
  coefficients
  $A_{m;j_1,\ldots,j_{p_{\kappa_{l}}}}^{l;\kappa_1,\ldots,\kappa_{l-1}}$,
  where $l$ denotes the layer depth, $\kappa_1,\ldots,\kappa_{l-1}$ denote the
  indices of the logical degrees of freedom starting from the top node and down
  to a particular node (path from the top down to a given node), $m$ indicates
  the different vectors within this node, and finally
  $j_1,\ldots,j_{p_{\kappa_{l}}}$ are the
  tensor indices of each particular array of coefficients within the node.

  % Node diagrams (mode combination, etc.)
  Some ML diagrams are presented later when discussing the various applications
  (see Figs.~\ref{fig:tree6},\ref{fig:tree18},\ref{fig:treepyr}).
  The lines connecting one node with its descendant nodes in such
  diagrams represent the tensor indices
  $j_1,\ldots,j_{p_{\kappa_1,\ldots,\kappa_l}}$, one line per index, and the
  numbers at the side of each line refer to the maximum possible value of the
  corresponding index. Each node is uniquely described by the values of its
  label $(\zn)$.

  \subsection{ML-MCTDH equations of motion and recursive
                       implementation}

  %%%%%% EOMs %%%%%%
  The ML-MCTDH equations of motion (EOM) have been derived and discussed by
  Wang and Thoss~\cite{wan03:1289} and Manthe~\cite{man08:164116}. In
  Ref.~\cite{man08:164116} a fully general derivation of the ML-MCTDH EOM for
  arbitrary layering schemes is provided, together with an algorithm for the
  recursive evaluation of all intermediate quantities entering the EOM. We
  reproduce here the EOM and hint the key elements of the recursive
  algorithm for the sake of completeness. Later we discuss some specific
  aspects of our implementation.

  The ML-MCTDH EOM have a very similar structure to the usual MCTDH equations,
  and for the top layer coefficients they are identical to the MCTDH ones
  \begin{equation} \label{eq:eom1}
    i \frac{\partial A_I^1}{\partial t} = \sum_J
    \langle \Phi_I^1 | \hat{H} | \Phi_J^1 \rangle A_J^1,
  \end{equation}
  where the top layer configurations
  $\Phi_J^1 = \varphi_{j_1}^{(1;1)}(Q_1^1,t)\cdot \ldots \cdot
  \varphi_{j_p}^{(1;p)}(Q_p^1,t)$,
  are defined as direct products of SPFs and the multi-index
  $J=j_1,\ldots,j_p$ has been implicitly introduced.
  The EOM for the propagation of the SPFs are formally the same for all layers
  \begin{equation} \label{eq:eom2}
    i \frac{\partial \varphi_n^{z,\kappa_l}}{\partial t} =
    (1-\hat{P}_{\kappa_l}^z) \sum_{j,m} (\rho^{z,\kappa_l})_{nj}^{-1} \cdot
    \langle \hat{H} \rangle_{jm}^{z,\kappa_l} \varphi_m^{z,\kappa_l},
  \end{equation}
  where $\hat{P}_{\kappa_l}^z=\sum_j
  |\varphi_j^{z,\kappa_l}\rangle\langle\varphi_j^{z,\kappa_l}|$ is
  the projector onto the space spanned by the $\varphi_j^{z,\kappa_l}$ SPFs,
  $\rho^{z,\kappa_l}$ is a density matrix and $\langle \hat{H}
  \rangle^{z,\kappa_l}$ is a matrix of mean-field operators acting on the
  $\varphi_j^{z,\kappa_l}$ functions. The symbol $z$ is a shorthand for the
  indices specifying a node, $z=\zn$, and for further reference we introduce
  $z\!-\!1=\znm$, and similar for $z\!+\!1$. In its form above, the EOM for
  the SPFs look identical to the usual EOM for the SPFs in the usual
  MCTDH~\cite{bec00:1}. Only the computation of the density matrices and
  mean-fields entering the EOM is now more involved than in a single-layer
  MCTDH scheme.

  %%%%%% Recursive algorithm %%%%%%
  Before describing how in practice the recursive calculation of mean-fields
  and density matrices is performed, we first restrict ourselves to the case in
  which the Hamiltonian is given by sums of products of operators acting on the
  primitive (but possibly combined) degrees of freedom $Q_\kappa^{\text{prim}}$
  \begin{equation} \label{eq:hamil1}
    \hat{H} = \sum_{r=1}^s c_r \hat{H}_r =
              \sum_{r=1}^s c_r \prod_{\kappa=1}^f \hat{h}_r^{(\mathrm{prim};\kappa)}.
  \end{equation}
  This form of the operator is necessary to avoid the computation of high
  dimensional integrals, and has been thoughtfully discussed in the MCTDH
  literature. Terms which usually do not follow the product form are, e.g.
  general potential energy surfaces. There exist however algorithms to bring
  general potentials to product
  form~\cite{jae96:7974,jae98:3772,lat00:1253,lat00:1324}.
  Furthermore, MCTDH and ML-MCTDH formulations based on time-dependent grids
  that by-pass this restriction (so-called CDVR approach) have also been
  provided~\cite{man96:6989,man08:164116}, but here we always will assume
  Hamiltonians in product form, Eq.~(\ref{eq:hamil1}).

  In order to discuss the practical computation of the mean-field elements
  $\langle \hat{H} \rangle_{jm}^{l;\kappa_1,\ldots,\kappa_l}$ appearing in
  Eq.~(\ref{eq:eom2}), we center our attention on a particular node $z=(\zn)$.
  This node relates to the logical coordinate
  $Q_{\kappa_{l-1}}^{l-1;\kappa_1,\ldots,\kappa_{l-2}}$, which can be further
  decomposed into a combination of logical coordinates
  \begin{equation} \label{eq:coords}
  Q_{\kappa_{l-1}}^{z\!-\!1}=
      \{Q_{1}^{z},\ldots,
        Q_{p_{\kappa_{l}}}^{z}\},
  \end{equation}
  and holds vectors of coefficients denoted by
  $A_{m;j_1,\ldots,j_{p_{\kappa_l}}}^{z}$.
  These coefficients are expansion coefficients of SPFs in layer $l\!-\!1$
  expanded in terms of SPFs in layer $l$. This is better seen by extending
  Eqs.~(\ref{eq:ansatz3},\ref{eq:ansatz5}) to the fully general multilayer case
  \begin{subequations} \label{eq:phiml}
  \begin{eqnarray}
     \varphi_m^{\Zml}
       (Q_{\kappa_{l-1}}^{z\!-\!1}) & = &
  \sum_{j_1}^{n_1} \cdots \sum_{j_{p_{\kappa_{l}}}}^{n_{\kappa_{l}}}
    A_{m;j_1,\ldots,j_{p_{\kappa_l}}}^{z}
    \prod_{\kappa_{l}=1}^{p_{\kappa_{l}}}
    \varphi_{j_{\kappa_{l}}}^{z,\kappa_{l}}
       (Q_{\kappa_{l}}^{z})  \\
     & = &
  \sum_{J} A_{m;J}^{z} \cdot
    \Phi_{J}^{z}
       (Q_{\kappa_{l-1}}^{z\!-\!1}),
  \end{eqnarray}
  \end{subequations}
  where the configurations $\Phi_J^z$ are the multilayer generalization
  of the configurations introduced in Eq.~(\ref{eq:eom1}).
  The operator summands in Eq.~(\ref{eq:hamil1}) can be written as
  \begin{subequations} \label{eq:hamil2}
  \begin{eqnarray}
    \hat{H}_r & = &  \hat{\mathcal{H}}_r^\z \cdot \hat{h}_r^\z \\
              & = &  \hat{\mathcal{H}}_r^\z \cdot
         \prod_{\kappa_{l+1}=1}^{p_{\kappa_{l+1}}} \hat{h}_r^{\zll},
  \end{eqnarray}
  \end{subequations}
  where $\hat{h}_r^\z$ acts on the logical coordinate
  $Q_{\kappa_l}^{l;\kappa_1,\ldots,\kappa_{l-1}}$,
  while $\hat{\mathcal{H}}_r^\z$ acts on the corresponding remaining space.
  As seen from Eq.~(\ref{eq:hamil2}b), $\hat{h}_r^\z$ can as well
  be written as a direct product of operators $\hat{h}_r^{\zll}$, which act on
  logical coordinates $Q_{\kappa_{l+1}}^{l+1;\kappa_1,\ldots,\kappa_{l}}$.
  In fact, both operators $\hat{\mathcal{H}}_r^\z$ and $\hat{h}_r^\z$
  can be broken down to a simple product of operators which act on one
  primitive (but possibly combined) coordinate only (see Eq.~(\ref{eq:hamil1})).
  The matrix elements of the mean-field operators at the right-hand-side of
  Eq.~\ref{eq:eom2} for one particular summand $\hat{H}_r$ are given by
  \begin{equation} \label{eq:meanf1}
    \langle \hat{H}_r \rangle_{jm,IJ}^{\Zl} =
    \langle \Phi_{I}^{z\!+\!1} |
       \langle \hat{H}_r \rangle_{jm}^{\Zl}
    |       \Phi_{J}^{z\!+\!1} \rangle,
  \end{equation}
  which is readily seen if the SPFs in Eq.~(\ref{eq:eom2}) are written in their
  explicit form given in Eq.~(\ref{eq:phiml}b).  These are the most cumbersome
  quantities to compute within the ML-MCTDH scheme.
  By substituting Eqs.~(\ref{eq:hamil2}a,b) into
  Eq.~(\ref{eq:meanf1}) one arrives at
  \begin{subequations} \label{eq:meanf2}
  \begin{eqnarray}
     \langle \hat{H}_r \rangle_{jm,IJ}^\z  & = &
        \langle \hat{\mathcal{H}}_r^\z \rangle_{jm}^\z \cdot
        \left[\hat{h}_r^\z\right]_{IJ} \\
     & = &
        \langle \hat{\mathcal{H}}_r^\z \rangle_{jm}^\z \cdot
        \prod_{\kappa_{l+1}=1}^{p_{\kappa_{l+1}}}
            \left[ \hat{h}_r^{\zll} \right]_{i_{\kappa_{l+1},j_{\kappa_{l+1}}}}
  \end{eqnarray}
  \end{subequations}
  with
  \begin{subequations} \label{eq:meanf3}
  \begin{eqnarray}
     \left[\hat{h}_r^\Zl\right]_{IJ} & = &
      \langle \Phi_{I}^{z\!+\!1} |
         \hat{h}_r^{\Zl}
    |       \Phi_{J}^{z\!+\!1} \rangle  \\
   \left[ \hat{h}_r^{\Zll} \right]_{i_{\kappa_{l+1},j_{\kappa_{l+1}}}} & = &
      \langle \varphi_{i_{\kappa_{l+1}}}^{\Zll} |
         \hat{h}_r^{\Zll}
    |       \varphi_{j_{\kappa_{l+1}}}^{\Zll} \rangle
  \end{eqnarray}
  \end{subequations}
  (The mean-field tensor $\langle \hat{\mathcal{H}}_r^\Zl \rangle_{jm}^\Zl $
  will be discussed below, Eq.~(\ref{eq:meanfgen})).
  It remains therefore to be seen, how the quantities appearing in the
  right-hand-side of Eq.~(\ref{eq:meanf2}b) can be recursively computed.

  %%% h-matrices %%%
  The expression for the $h$-matrices is found by starting
  from Eq.~(\ref{eq:meanf3}b) and explicitly writing the SPFs in layer
  $l+1$ in terms of those in layer $l+2$
  \begin{eqnarray} \label{eq:tmat}
   \nonumber
   \left[\hat{h}_{r}^{\Zll}\right]_{a_{\kappa_{l+1}},b_{\kappa_{l+1}}}
   & = & \langle \varphi_{a_{\kappa_{l+1}}}^{\Zll} |
                     \hat{h}_r^\Zll
         | \varphi_{b_{\kappa_{l+1}}}^{\Zll}  \rangle      \\
   & = & \langle \varphi_{a_{\kappa_{l+1}}}^{\Zll} |
    \displaystyle\prod_{\kappa_{l+2}=1}^{p_{\kappa_{l+2}}} \hat{h}_r^{\Zlll}
         | \varphi_{b_{\kappa_{l+1}}}^{\Zll}  \rangle      \\
   \nonumber
   & = & \displaystyle\prod_{\kappa_{l+2}=1}^{p_{\kappa_{l+2}}}
       \displaystyle\sum_{J^{\kappa_{l+2}}}
       \displaystyle\sum_{i,j}
     A_{a_{k_{l+1}};J_{i}^{\kappa_{l+2}}}^{*\;z\!+\!2}
         \left[ \hat{h}_r^{\Zlll} \right]_{i,j}
     A_{b_{k_{l+1}};J_{j}^{\kappa_{l+2}}}^{z\!+\!2}.
  \end{eqnarray}
  Here multi-indices $J^\kappa$ and $J_m^\kappa$ have been introduced to make
  the notation more compact. The former contains all indices except for the
  $\kappa$-th one, while the latter has index $\kappa$ set to $m$.
  Eq.~(\ref{eq:tmat}) tells us that the computation of the
  matrix elements of $\hat{h}_r^{\Zll}$ requires the previous knowledge of
  the matrix elements of $\hat{h}_r^{\Zlll}$.
  Therefore, one starts at the lowest layer with the matrix elements of $\hat{H}$
  with respect to the underlying time-independent basis
  \begin{equation} \label{eq:tmatprim}
    \left[ \hat{h}_r^{l_{\mathrm{max}};\kappa_1,\ldots,\kappa_{l_{\mathrm{max}}}}
    \right]_{i_{\kappa},j_{\kappa}} =
    \left[ \hat{h}_r^{\mathrm{prim};\kappa} \right]_{i_{\kappa},j_{\kappa}}
      = \langle \chi_{i_\kappa}^{(\kappa)} |
                     \hat{h}_r^{\mathrm{prim};\kappa}  |
                \chi_{j_\kappa}^{(\kappa)} \rangle
  \end{equation}
  where $l_{\mathrm{max}}$ denotes the maximal layer, i.e. $l_{\mathrm{max}}=2$ for
  normal MCTDH,
  and proceeds bottom up through the tree structure building the matrix
  representations of the $\hat{h}_r^\zll$ operators, which will subsequently
  be used to build the matrix elements of $\hat{h}_r^\z$ operators, and so on.

  %%% mean-field specific part %%%
  The mean-field matrix elements are computed in practice in a similar way to
  Eq.~(\ref{eq:tmat}). For the top layer, the mean-field elements are given as
  \begin{equation} \label{eq:meanftop}
    \langle \hat{\mathcal{H}}_r^{1;\kappa_1} \rangle_{m,n}^{1;\kappa_1} =
    \displaystyle\prod_{\nu\neq\kappa_1}^p
    \displaystyle\sum_{J^{\kappa_{1},\nu}}
    \displaystyle\sum_{i,j}
         A_{1;J_{m,i}^{\kappa_{1},\nu}}^{*\;1}
       \left[ \hat{h}_r^{0;\nu} \right]_{i,j}
         A_{1;J_{n,j}^{\kappa_{1,\nu}}}^{1},
  \end{equation}
  where the composite indices $J^{\kappa,\nu}$ and $J_{m,n}^{\kappa,\nu}$ are
  defined similarly as $J^\kappa$ and $J_m^\kappa$.  Eq.~(\ref{eq:meanftop})
  corresponds to the normal way of calculating the mean-fields in standard
  MCTDH, and is an alternate way of writing, e.g. Eq.~(67) in
  Ref.~\cite{bec00:1}.
  For a given deeper layer, the calculation involves the mean-fields of the above
  layer and is given by~\cite{man08:164116}
  \begin{eqnarray} \label{eq:meanfgen}
    \nonumber
    \langle \hat{\mathcal{H}}_r^{\Zl} \rangle_{m,n}^{\Zl} =
    \displaystyle\sum_{a,b}
    \langle \hat{\mathcal{H}}_r^{\Zml}
                       \rangle_{a,b}^{\Zml} \\
    \times \displaystyle\prod_{\nu\neq\kappa_l}^{p_{\kappa_l}}
    \displaystyle\sum_{J^{\kappa_{l},\nu}}
    \displaystyle\sum_{i,j}
         A_{a;J_{m,i}^{\kappa_{1},\nu}}^{*\;z}
       \left[ \hat{h}_r^{z\!-\!1,\nu} \right]_{i,j}
         A_{b;J_{n,j}^{\kappa_{1},\nu}}^z \; ,
  \end{eqnarray}
  where $a$ and $b$ run over the $(l-1)$th layer SPFs, which are contained
  in the $l$th layer SPF $m$ and $n$, respectively.
  Therefore, the mean-fields can be computed if the mean-fields of the layers
  above, as well as the matrix elements of the $\hat{h}$ operators, are known. The
  density matrices are computed in a very similar way to Eq.~(\ref{eq:meanfgen}),
  but application of the $\hat{h}$ matrices to $A$ vectors on the right side of
  Eq.~(\ref{eq:meanfgen}) disappears.
  Eqs.~(\ref{eq:meanftop}-\ref{eq:meanfgen}) are obtained after introducing the
  so called single-hole functions, which are defined as the variation of the total
  wavefunction with respect to a particular SPF. We leave this more specialized step
  out of this discussion and refer the interested reader, e.g. to Refs. \cite{bec00:1}
  or \cite{man08:164116}.

  The recursive algorithm proceeds then as follows: 1) Starting from the bottom
  of the tree and proceeding up, the $h$-matrix elements are evaluated after
  Eq.~(\ref{eq:tmat}). 2) Starting from the top of the tree and proceeding down
  the mean-fields and density matrices are computed as in
  Eq.~\ref{eq:meanfgen}.  3) The right-hand-side of the EOMs,
  Eqs.~(\ref{eq:eom1},\ref{eq:eom2}), is evaluated.

  \subsection{Separable part of the Hamiltonian across layers}

  The efficiency of the ML-MCTDH calculations is increased if one takes
  advantage of the terms $\hat{h}_r^{(\mathrm{prim};\kappa)}$ in
  Eq.~(\ref{eq:hamil1}) that are unit operators.  The kinetic energy part of
  the Hamiltonian contains usually many unit terms.  Also common models such as
  Henon-Heiles, vibronic Coupling, system-bath, etc. have as well many unit
  terms in the potential energy part of the operator.

  The obvious way to obtain computational savings is to flag all unit terms as
  such, so that they do not have to be stored as unit matrices and they do not
  have to be explicitly multiplied. This is done from the bottom of the tree and
  proceeding recursively upwards. If all
  $\hat{h}_r^{\Zll}\equiv\hat{1}$, then $\hat{h}_r^{\Zl}$ is flagged as
  $\hat{1}$. Our implementation keeps track of all
  unit $h$-terms at all layers, which saves resources in the recursive
  construction of the $h$-matrices in Eq.~(\ref{eq:tmat}), in the mean-fields
  construction in Eq.~(\ref{eq:meanfgen}) and in the equations of motion.

  Once the unit operators at all layers have been flagged as such, it is
  possible to identify non-correlated terms for a particular node, i.e. terms
  that decompose as
  \begin{equation} \label{eq:hamil-sep}
    \hat{H}_r = \mathbf{1}^{\Zl} \cdot
         \hat{h}_r^{\Zl},
  \end{equation}
  such that the non-unit parts of $H_r$ operate only on the combined coordinate
  $Q_{\kappa_{l}}^{z}$.  The step of identifying
  separable terms is conveniently done from the top of the tree and proceeding
  downwards.  To illustrate this let us refer to a simple 4D example
  and the operator summand
  \begin{equation} \label{eq:op4D}
     \hat{H}_r =
        \hat{h}_r^{\mathrm{prim};1} \cdot
        \hat{1}_r^{\mathrm{prim};2} \cdot
        \hat{h}_r^{\mathrm{prim};3} \cdot
        \hat{h}_r^{\mathrm{prim};4}.
  \end{equation}
  The considered tree structure has four primitive nodes
  $(3;1,1)$, $(3;1,2)$, $(3;2,1)$ and $(3;2,2)$
  with $h$-terms
  $\hat{h}_r^{3;1,1}$,
  $\hat{1}_r^{3;1,2}$,
  $\hat{h}_r^{3;2,1}$ and
  $\hat{h}_r^{3;2,2}$,
  which are the matrix representation of the primitive product terms in
  Eq.~(\ref{eq:op4D}).
  The two nodes at the intermediate layer, namely $(2;1)$ and $(2;2)$, have
  therefore $h$-terms
  $\hat{h}_r^{2;1,1}$ and
  $\hat{1}_r^{2;1,2}$
  for node $(2;1)$ and $h$-terms
  $\hat{h}_r^{2;2,1}$ and
  $\hat{h}_r^{2;2,2}$
  for node $(2;2)$.
  The top node $(1;)$ has $h$-terms
  $\hat{h}_r^{1;1}$ and
  $\hat{h}_r^{1;2}$.
  Only in the case that both $\hat{h}_r^{2;1,1}$ and $\hat{1}_r^{2;1,2}$ would
  have been unit operators, would $\hat{h}_r^{1;1}$ have been marked as a unit
  operator as well.
  Now we turn to the identification of non-correlated terms starting from the
  top of the tree. $\hat{H}_r$ is marked as correlated in node $(1;)$ because both
  $\hat{h}_r^{1;1}$ and $\hat{h}_r^{1;2}$ are different from the unit operator.
  At first sight, the $\hat{H}_r$ term could be flagged as non-correlated in node
  $(2;1)$ because only one of the $h$-terms is different from unity. However,
  the parent node had this term marked as correlated, and therefore $\hat{H}_r$
  is marked as correlated in node $(2;1)$. The reason why this happens is that
  $\hat{H}_r$ acts also on node $(2;2)$, which is in another branch of the
  tree. It is then clear that a convenient way to spot situations of this kind
  is to proceed recursively from the top of the tree, as illustrated.
  Ultimately, the $\hat{H}_r$ term is marked correlated in node $(2;1)$ because
  the mean-field matrix elements
  $\langle \hat{\mathcal{H}}_r^{2;1,1} \rangle_{m,n}^{2;1,1}$
  and
  $\langle \hat{\mathcal{H}}_r^{2;1,2} \rangle_{m,n}^{2;1,2}$ correlate the
  motion of SPFs $\varphi_j^{2;1,1}$ and $\varphi_j^{2;1,2}$ with the dynamics
  proceeding in node $(2;2)$ and below. The coupling to the dynamics in the
  other branch arises from the relation of the mean-field elements just
  mentioned to the mean-fields $\langle \hat{\mathcal{H}}_r^{1;1}
  \rangle_{m,n}^{1;1}$ in the top node, as shown in Eq.~(\ref{eq:meanfgen}).

  Once the non-correlated terms have been identified for each node, one can
  exploit, in the same way as in standard MCTDH, that for them
  the mean-field matrix elements become identical to the
  elements of the density matrix
  \begin{equation} \label{eq:unitMF}
    \langle \hat{\mathbf{1}}^\z \rangle_{mn}^{\z} \equiv (\rho^\z)_{mn}.
  \end{equation}
  Hence, for a particular node with $\hat{H}_r$ terms $1$ to $s$
  non-correlated, and terms $s+1$ to $t$ correlated, Eq.~(\ref{eq:eom2}) can be
  rewritten as
  \begin{equation} \label{eq:eom2sep}
    i \frac{\partial \varphi_n^{z,\kappa_l}}{\partial t} =
    \left(1-\hat{P}_{\kappa_l}^z\right)
    \left(
    \sum_{q=1}^s c_q\cdot\hat{h}_q^{z,\kappa_l} +
    \sum_{j,m} (\rho^{z,\kappa_l})_{nj}^{-1} \cdot
    \sum_{r=s+1}^t
        c_r\cdot\langle\hat{\mathcal{H}}_r^{z,\kappa_l}\rangle_{jm}^{z,\kappa_l}\cdot
    \hat{h}_r^{z,\kappa_l}
    \right)\varphi_m^{z,\kappa_l},
  \end{equation}
  which is the EOM for the SPFs on which our implementation is based.
  The EOM for the coefficients follows directly from this equation
  and Eq.~(\ref{eq:phiml})
  \begin{equation} \label{eq:eom2coef}
    \frac{\partial A_{m;J}^{z} }{\partial t} = \langle  \Phi_{J}^{z} |
    \frac{\partial \varphi_n^{\Zml}}{\partial t} \rangle \; .
  \end{equation}
  There is no contribution from $\partial{\Phi_{J}^{z}}/\partial t$ because,
  due to the projector, the SPFs are orthogonal to their time derivatives.
  Hence one may write compactly
  \begin{equation} \label{eq:eom2coef2}
    \frac{\partial A_{n;J}^z }{\partial t} =  \sum_{K}
   \big(\delta_{JK} - \sum_j A_{j;J}^z A_{j;K}^{* z}\big)\,
    \sum_{m,L} M^z_{n,m;KL}\, A_{m;L}^z \; ,
  \end{equation}
  where a detailed expression for the matrix $M^z_{nm;JL}$ follows from
  Eqs.~(\ref{eq:eom2sep}, \ref{eq:eom2coef}, \ref{eq:phiml}).
 
  As a final comment, it is worth mentioning that in large systems with groups
  of coordinates strongly correlated among them but weakly
  correlated to other groups, i.e. with some sort of locality, the
  identification of the correlated terms node by node may become very
  important. In such cases there will be many correlated terms in nodes at
  lower layers, but less of them as one proceeds up to the top of the tree.
  Such an example is given by be the HH model to be introduced below,
  where each harmonic oscillator is coupled only to its next neighbors.  For
  similar reasons, when dealing with system-bath Hamiltonians with ML-MCTDH, it
  may be advantageous in most cases to separate system and bath already at the
  top layer.

\section{Results and Discussion}  \label{sec:results}

  The ML-MCTDH algorithm and implementation described above are applied to two
  paradigmatic systems. On the one hand we simulate the HH model
  for various dimensionalities, and on the other hand we test our
  implementation on the pyrazine system using the second-order
  vibronic-coupling Hamiltonian of Ref.~\cite{raa99:936}.

  \subsection{Henon-Heiles} \label{sec:hh}

  The HH Hamiltonian
  \begin{equation} \label{eq:hh_model}
    \hat{H} = \frac{\omega}{2} \sum_{\kappa=1}^f
              \left(
                - \frac{\partial^2}{\partial q_\kappa^2}
                + q_\kappa^2
              \right)
              +
              \lambda \sum_{\kappa=1}^{f-1}
              \left(
                 q_{\kappa}^2\; q_{\kappa+1} -\frac{1}{3} q_{\kappa+1}^3,
              \right)
  \end{equation}
  written here in dimensionless units, offers a convenient playground to
  investigate the convergence properties and performance of ML-MCTDH.
  The HH model was used in Ref.~\cite{nes02:10499} to benchmark the MCTDH
  method and the same Hamiltonian with similar parameters is used here
  for the bechmarking of ML-MCTDH.
  For all simulations we set $\omega=1$. The degree of anharmonicity and
  coupling are controlled by the single parameter $\lambda$, for which different
  values are chosen. The initial amount
  of energy in the system is easily controlled by the position at which
  the initial wavepacket is centered for each degree of freedom.  The
  initial wavepacket is always taken as a Hartree product of Gaussian functions
  of width corresponding to the ground vibrational state of the harmonic part
  of the Hamiltonian. Initially, all momenta are set to zero, $\langle
  p_\kappa\rangle=0$, and similarly all positions, except for a few coordinates
  which are displaced by two length units with respect to $q_\kappa=0$.  The HH
  Hamiltonian in Eq.~(\ref{eq:hh_model}) has dissociative channels due to the
  cubic terms, which are accessible in the energy range of the reported
  simulations. To avoid reflections from the grid edges, the Hamiltonian is
  augmented with a complex absorbing potential (CAP) as $H=T+V-iW$. Further
  details on the CAP used are found in Ref.~\cite{nes02:10499}. A sine-DVR (for
  discrete variable representation) primitive basis with points between $-9$
  and $7$ length units and $N=24$ grid points is used for all coordinates
  throughout the HH simulations. All propagations on HH models are carried on
  up to 30 time units, which roughly corresponds to five oscillation periods
  $T=2\pi/\omega$ in the harmonic limit.

  % 6D:
  \subsubsection{6D simulations}
  \placefig{fig:tree6}
  We first concentrate on a 6D HH model.
  Calculations with $\lambda=\lambda_0$ and $\lambda=2\lambda_0$ are performed,
  for $\lambda_0=0.111803$. A value of $\lambda_0=0.111803$
  allows to compare with other works that use the same
  Hamiltonian~\cite{nes02:10499}. For the
  reported 6D calculations, the initial wavepacket is centered at $q_\kappa=2$
  for coordinates $q_2$, $q_4$ and $q_6$.  Standard MCTDH (2-layer) and
  ML-MCTDH calculations using a 3-layer scheme are compared, and the trees
  representing the MCTDH and ML-MCTDH wavefunctions are displayed in
  Figs.~\ref{fig:tree6}a and \ref{fig:tree6}b, respectively. In the MCTDH case
  every two primitive DOF are combined, and $N_2$ equals
  the number of DVR grid points (primitive basis functions) for each DOF. In
  the ML-MCTDH wavefunction an extra layer of TD coefficients is introduced,
  effectively representing the 2D SPFs of the combined modes in the original
  MCTDH wavefunction by a new multiconfigurational expansion. It is worth
  noting that in the case of having $N_2=N_3$ for the ML-MCTDH wavefunction,
  the lowest layer is treated exactly and both MCTDH with mode combination and
  ML-MCTDH provide numerically identical results (for the same number of SPFs
  $N_1$ at the top layer).

  The different propagations are compared by inspecting the autocorrelation
  function obtained as $a(t) = \langle \Psi^*(t/2)|\Psi(t/2)\rangle$.
  (This trick~\cite{bec00:1} yields an autocorrelation function which is
  twice as long as the propagation).
  The autocorrelation at long times is difficult to converge, since the
  (ML-)MCTDH wavefunction accumulates error as time increases, providing an
  adequate quantity to compare the different propagations. Conversely, the
  spectra corresponding to the autocorrelation functions have the error
  averaged over the whole energy range, and therefore are not such a good
  quantity to look at for a stringent comparison.

\begin{figure}[h!]
      \begin{center}
        \includegraphics[width=8.5cm]{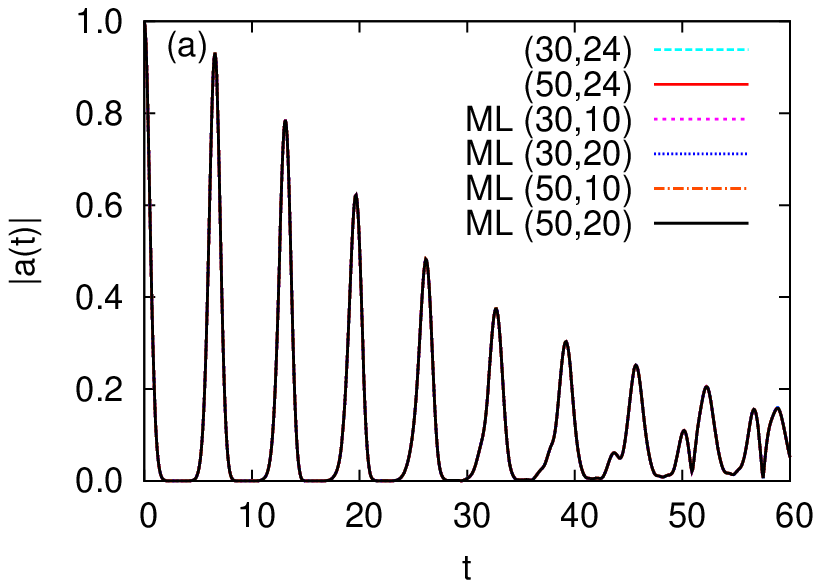}
        \includegraphics[width=8.5cm]{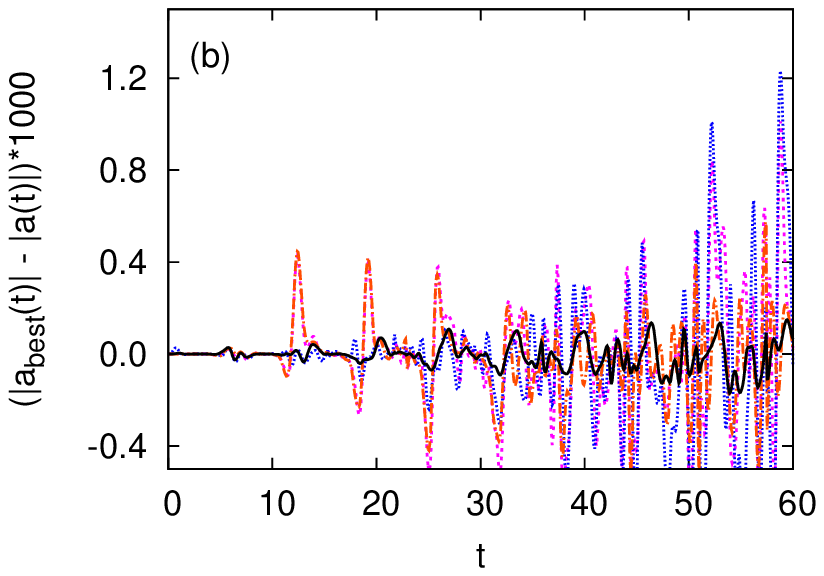}
      \end{center}
     \caption{\figfoot}
     \label{fig:auto6D1}
    \end{figure}

\begin{figure}[h!]
      \begin{center}
        \includegraphics[width=8.5cm]{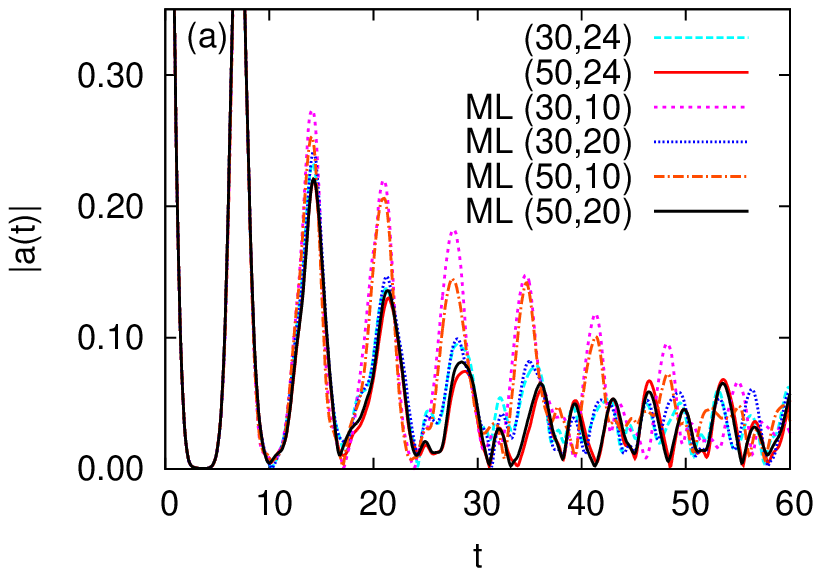}
        \includegraphics[width=8.5cm]{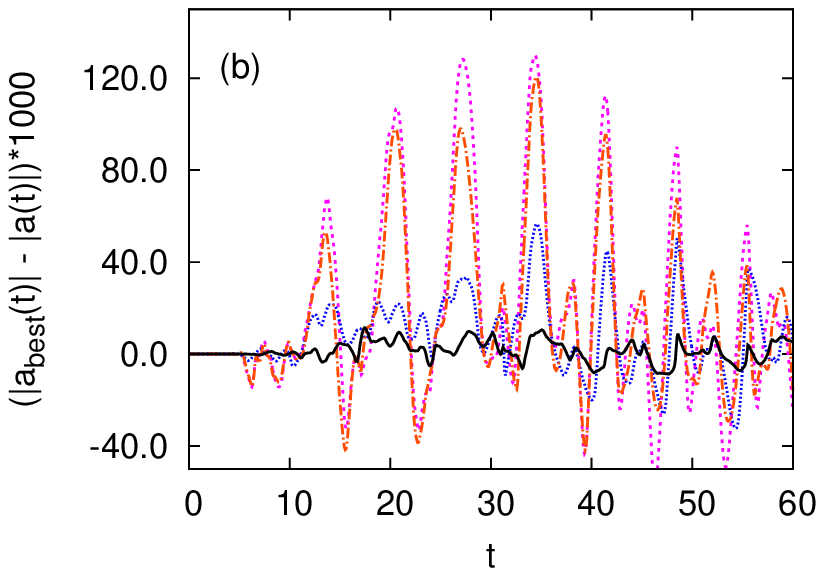}
      \end{center}
     \caption{\figfoot}
     \label{fig:auto6D2}
    \end{figure}

  \placefig{fig:auto6D1}
  \placefig{fig:auto6D2}
  Results for the low ($\lambda=\lambda_0$) and high ($\lambda=2\lambda_0$)
  coupling regimes are shown in Figs.~\ref{fig:auto6D1} and \ref{fig:auto6D2},
  respectively, where the absolute value of $a(t)$ is displayed.  In both
  Figs.~\ref{fig:auto6D1} and \ref{fig:auto6D2}, the numbers in parenthesis
  correspond
  to $(N_1,N_2)$ in the trees in Fig.~\ref{fig:tree6}. For MCTDH $N_2$ equals
  the number of grid points and is given only for completeness.
  In the case of low coupling (Fig.~\ref{fig:auto6D1}), all calculations
  yield almost indistinguishable autocorrelation functions.
  Fig.~\ref{fig:auto6D1}b displays the difference between autocorrelation
  functions of the ML-MCTDH calculations and the best converged (50,24) MCTDH
  calculation.  The (50,20) ML-MCTDH calculation is very close to the reference
  MCTDH result. The calculations that diverge earlier with respect to the reference
  result are the (30,10) and (50,10) ones due a poorer convergence of the
  lower layer, while the ML-MCTDH (30,20) calculation starts to deviate from the
  reference result after the system has completed five to six oscillation
  cycles.

  After doubling the strength of the coupling parameter, the situation becomes
  more complex. In Fig.~\ref{fig:auto6D2}a one sees that after two oscillation
  periods the autocorrelation functions of the different propagations start to
  diverge on an appreciable scale. Only the best MCTDH calculation (50,24), and
  the best ML-MCTDH (50,20) result in very close autocorrelations up to 60 time
  units.  Fig.~\ref{fig:auto6D2}b shows the difference between the
  ML-MCTDH results and the (50,24) MCTDH calculation.  The
  two worst calculations are clearly the ones with $N_2=10$. In this case, the
  correlation at the lowest layer, i.e.  between DOF $(q_1,q_2)$, $(q_3,q_4)$
  and $(q_4,q_5)$ is not well represented.  It is important to note that the
  (50,10) calculation does {\em not} offer an improvement over the (30,10) one.
  The missing correlation within the logical coordinates in the lowest layer
  cannot be regained by having more SPFs in higher layers.
  Similarly as in standard MCTDH calculations, the eigenvalues of the density
  matrices at each layer, the natural populations (NP), provide an indication
  of the degree of convergence of a calculation. As a rule of thumb, the
  smallest NP should be of the order of $10^{-5}$ to provide good results in
  many cases, although this can vary depending on what quantity one is trying
  to compute.  For the (50,10) and (50,20) ML-MCTDH calculations
  at $t=60$, the smallest NP of the three logical
  coordinates at the top layer are $(1.4\cdot 10^{-5}\,/\,1.3\cdot
  10^{-4}\,/\,9.3\cdot 10^{-5})$ and $(3.4\cdot 10^{-5}\,/\,1.8\cdot
  10^{-4}\,/\,1.3\cdot 10^{-4})$, respectively.
  Therefore, the convergence at the top layer is slightly worse for the (50,20)
  case than for the (50,10) case, even though by comparison to the standard
  MCTDH calculation it is clear that the (50,20) calculation provides better
  converged results than the (50,10).  Hence, by allowing more SPFs at the
  second layer, the evolution of the top layer SPFs becomes more complex since
  they have more variational freedom.

  As a remark to the 6D simulations, in this case standard MCTDH calculations
  are more efficient than ML-MCTDH ones. Only in the case of having a larger
  number of DVR points per degree of freedom resulting in very large 2D
  combined coordinates, would the introduction of a further layer (compare
  Figs.~\ref{fig:tree6}a and \ref{fig:tree6}b) constitute an advantage.
  Efficiency issues are discussed later in relation to the larger HH
  simulations and to pyrazine.

  \subsubsection{18D simulations}
\begin{figure}[h!]
      \begin{center}
        \includegraphics[width=8.5cm]{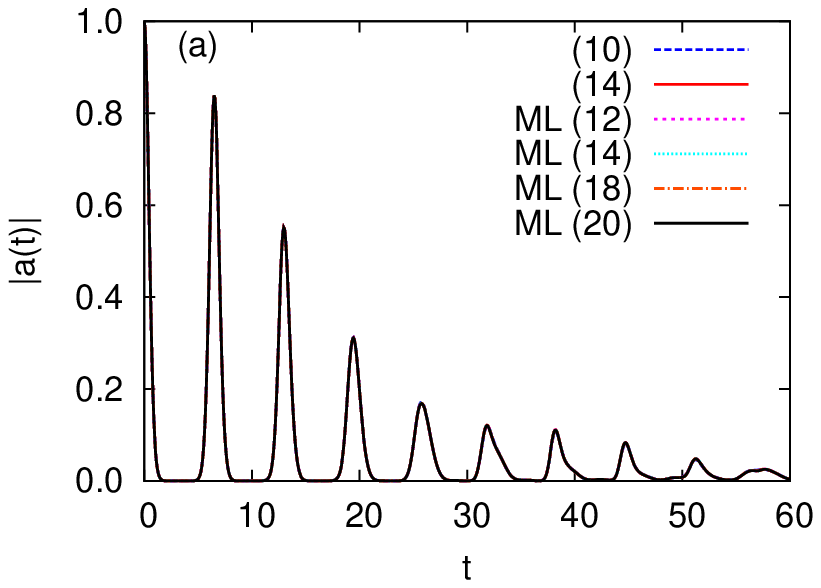}
        \includegraphics[width=8.5cm]{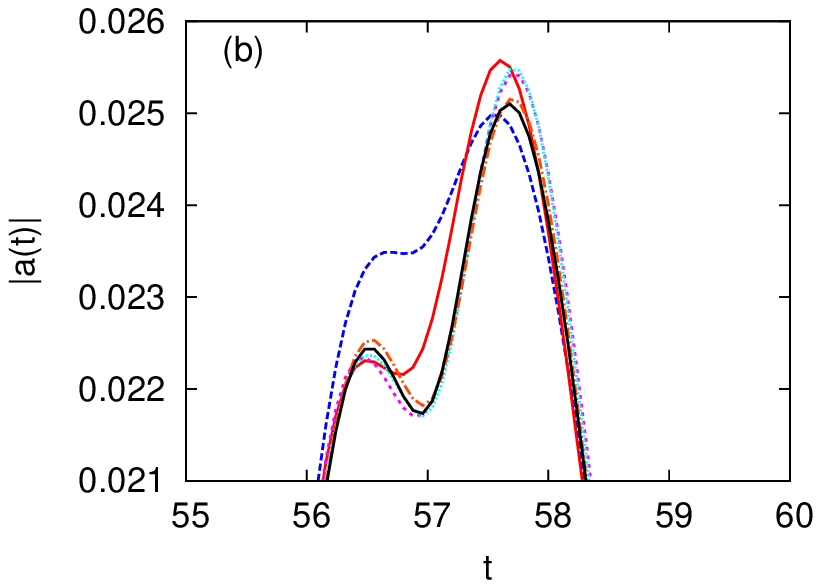}
      \end{center}
     \caption{\figfoot}
     \label{fig:auto18D}
    \end{figure}

  % 18D: compare MCTDH with ML, convergence,...
  \placefig{fig:tree18}
  Next we discuss simulations on a 18D HH model with
  $\lambda=\lambda_0$. The initial wavepacket is centered at $q_\kappa=2$ for
  coordinates $q_4$, $q_8$, $q_{12}$ and $q_{16}$, resulting in an initial
  energy of $15.807$. As in the 6D case, wavepackets are propagated
  up to 30 time units.
  For this model we run standard MCTDH calculations in which the 18 physical
  coordinates are grouped into 6 combined modes (logical coordinates) with
  three primitive DOFs each. The tree representation of the MCTDH wavefunction
  is given in Fig.~\ref{fig:tree18}a. There, $N_1$ is the number of SPFs per
  combined mode and $N_2=24$ is the number of DVR functions per DOF.  Two
  standard MCTDH calculations were performed, using $N_1=10$ and $N_1=14$ SPFs
  per combined mode, resulting in $10^6$ and $7.5\cdot10^6$
  configurations, respectively.
  ML-MCTDH calculations based on the tree structure in Fig.~\ref{fig:tree18}b
  and using the same Hamiltonian and initial conditions were performed as well.
  In the ML calculations various basis sizes were employed, always keeping
  $N_1$ equal to $N_2$. The autocorrelation functions resulting from the MCTDH
  and ML-MCTDH calculations are presented in Fig.~\ref{fig:auto18D}, where the
  numbers in the legend correspond to $N_1$ for MCTDH, and to $N_1$ and $N_2$
  for ML-MCTDH.
  At the scale shown in Fig.~\ref{fig:auto18D}a all simulations yield very
  similar results. Fig.~\ref{fig:auto18D}b shows the last revival of $|a(t)|$
  in more detail.  The worst result appears to be the standard MCTDH
  calculation with $N_1=10$. The various ML-MCTDH calculations approach the
  $(N_1,N_2) = 20$ result as the size of the basis increases, while the
  standard MCTDH with $N_1=14$ seems not to be yet be fully converged with
  respect to the trend of the ML-MCTDH simulations.

  \placefig{fig:auto18D}

  % 18 Strong interaction
  The set of simulations described above were repeated doubling coupling
  parameter \mbox{$\lambda=2\lambda_0$}. For this coupling strength the
  revivals of the autocorrelation function decay after a few oscillations.  The
  different simulations yield still similar autocorrelation functions
  (Fig.~\ref{fig:auto18DSI}a), but differences are now larger than in the small
  coupling parameter case, which can be seen by inspecting
  Fig.~\ref{fig:auto18DSI}b.  There, the autocorrelation function between 10
  and 30 time units is shown in detail, and after this time the autocorrelation
  has almost vanished.
  By looking at the autocorrelation functions of the various MCTDH and ML-MCTDH
  simulations one sees that full convergence with respect to the number of SPFs
  has not been completely reached, even at short propagation times. Full
  convergence of the 18D HH model with a coupling parameter of
  $\lambda=2\lambda_0$ is a very demanding task. The $N_1=14$ MCTDH simulation
  was run on 8 processors using shared-memory parallelization, for which a
  speed factor of about 3 can be expected~\footnote{MCTDH parallelizes rather
  poorly in the present case  because the Hamilton operator consist of only
  few terms. For other systems, e.g. the Zundel cation~\cite{ven09:034308},
  a much better parallelization performance has been
  reached~\cite{bri08:141,bri09:67,bri10:147}}.
  Such calculation required 71 hours of wall clock time on the 8 processors and
  would have taken about 200 hours on a single processor. About 5 GB of main
  memory were used for the wavepacket propagation.
  In contrast to this, the largest ML-MCTDH calculation with $(N_1,N_2)=20$
  required 136 hours of wall-clock time running on a {\em single} processor and
  the wavepacket propagation used less than 500 MB of main memory, while the
  $(N_1,N_2)=12$ ML-MCTDH calculation needed only 7 hours on a single
  processor. All calculations reported in this paper are performed on an
  quad-core Opteron, processor type 2384, 2.7 GHz.

  \placefig{fig:auto18DSI}

\begin{figure}[h!]
      \begin{center}
        \includegraphics[width=8.5cm]{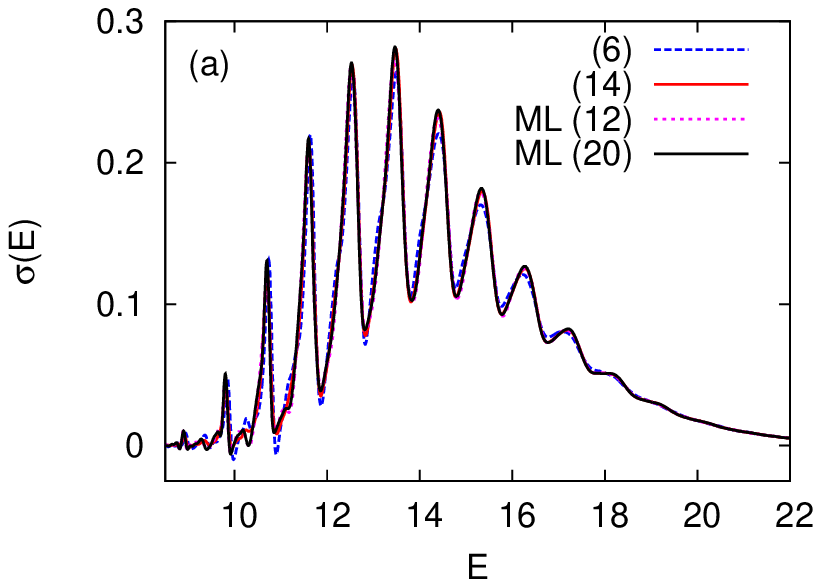}
        \includegraphics[width=8.5cm]{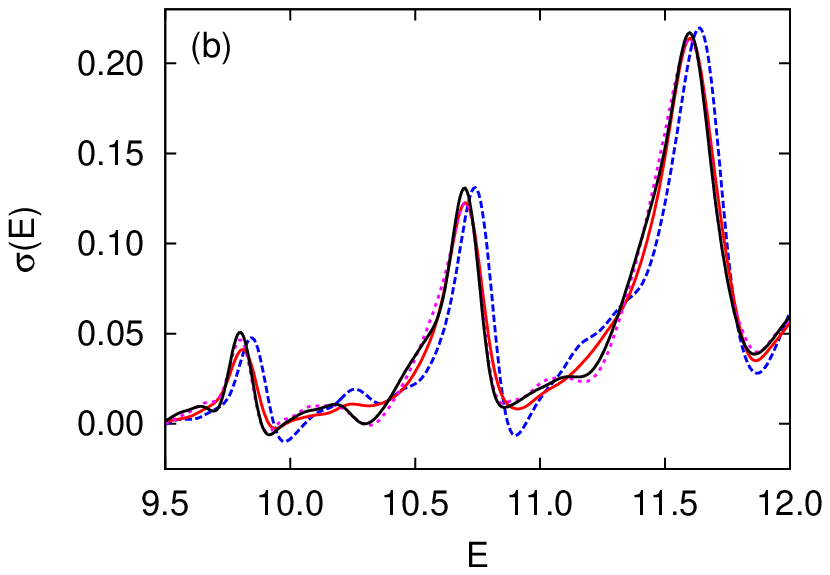}
      \end{center}
     \caption{\figfoot}
     \label{fig:spec18DSI}
    \end{figure}

  If we now look at the spectra of the propagated wavepackets obtained by
  Fourier Transform (FT) of the corresponding autocorrelation functions we see
  that they are very similar.  The spectra of two MCTDH calculations with
  $N_1=6$ and $N_1=14$ are shown in Fig.~\ref{fig:spec18DSI}. The $N_1=10$ case
  (not shown) yields a spectrum very similar to the larger MCTDH calculation.
  The spectra of two ML-MCTDH calculations are also presented in
  Fig.~\ref{fig:spec18DSI}, namely the smallest and largest calculations of the
  ML series.  The spectrum of the MCTDH calculation with $N_1=6$ yields
  somewhat different intensities in the higher energy range than the other
  simulations. When closely inspecting the lower energy range
  (Fig.~\ref{fig:spec18DSI}b) one finds a non-negligible energy shift of the
  spectrum of the small MCTDH calculation to higher energies. Interestingly,
  the $N_1=6$ standard MCTDH propagation required 17 hours running on 8
  parallel processors in the same conditions as discussed above. The number of
  SPFs for this simulation was chosen such that the cost would be comparable to
  that of the smaller $(N_1,N_2)=12$ ML-MCTDH calculation. The smaller ML-MCTDH
  calculation required 7 hours of wall-clock time on a single CPU running on
  the same hardware, and its spectrum is already very similar to the spectra of
  the larger MCTDH and ML-MCTDH calculations.

  \placefig{fig:spec18DSI}

  The size of the 18D HH model is such that standard MCTDH calculations can
  still be conducted to a good accuracy. Contrary to the 6D case however, for
  this dimensionality it already pays off to introduce a further layer
  in the calculation, and 3-layer ML calculations offer results of similar
  quality than the standard MCTDH ones with a noticeably smaller cost.

  \subsubsection{1458D simulations}

  Finally, we report ML-MCTDH simulations on a HH model with 1458 DOF and
  coupling constant $\lambda=\lambda_0$. The system is described in this case
  by a 7-layer wavefunction.  At the top level the coordinates are divided in
  three groups of 486 coordinates, which are subsequently divided again in
  three groups, and so on.  This is repeated until groups of two primitive DOF
  are reached in the seventh layer, which are then kept combined.
  For this deeply layered wavefunction two limiting cases are numerically
  investigated. In example (1), DOF $q_{486}$ and $q_{487}$ are
  displaced for $t=0$ such that $\langle q_\kappa \rangle=2$. Since the two
  degrees of freedom are adjacent, there is a coupling term in the Hamiltonian
  between them. However, $q_{486}$ and $q_{487}$ belong to different logical
  coordinates at all layers of the tree. They
  belong to logical coordinates $Q^1_1$ and $Q^1_2$ at the top layer and
  to coordinates $Q^{6;1,3,3,3,3}_3$ and $Q^{6;2,1,1,1,1}_1$ at the bottom
  layer, respectively.
  In example (2), DOF $q_{729}$ and $q_{730}$ are initially set at $\langle
  q_\kappa \rangle=2$. Again, these two DOF are coupled in the Hamiltonian, but
  in this case they belong to the same logical coordinate at all levels,
  starting from logical coordinate $Q^1_2$ at the top layer and down to the
  same combined mode $Q^{6;2,2,2,2,2}_2$ at the bottom layer.
  In both cases, decoupled reference calculations are also conducted in which
  the coupling term between DOF $q_{486}$ and $q_{487}$ for case (1) and
  between DOF $q_{729}$ and $q_{730}$ for case (2) are eliminated from the
  Hamiltonian.
  The simulated system corresponds to a long chain of coupled oscillators. In
  both cases (1) and (2), the initially displaced degrees of freedom are far
  from the ends of the chain.  Therefore the system dynamics in both cases
  should be very similar, since the extremes of the chain will play a role
  only at much longer times than simulated.

\begin{figure}[h!]
      \begin{center}
        \includegraphics[width=8.5cm]{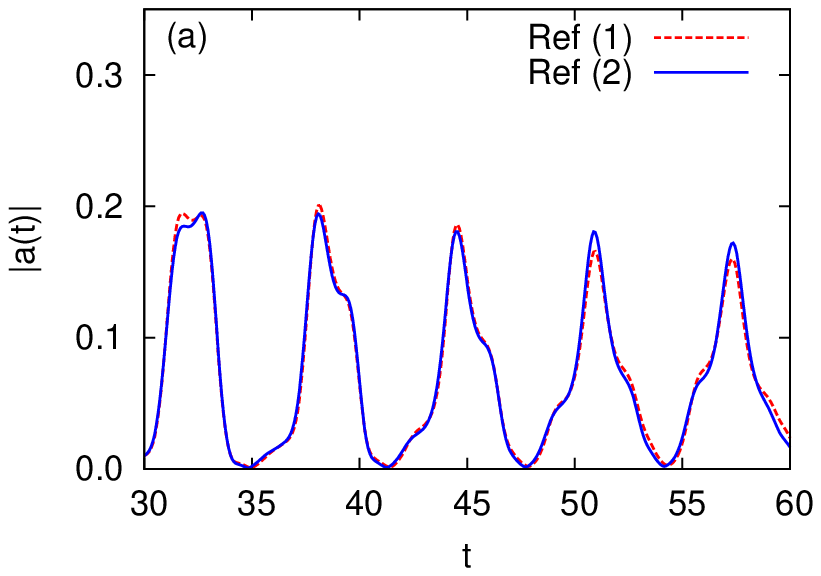}
        \includegraphics[width=8.5cm]{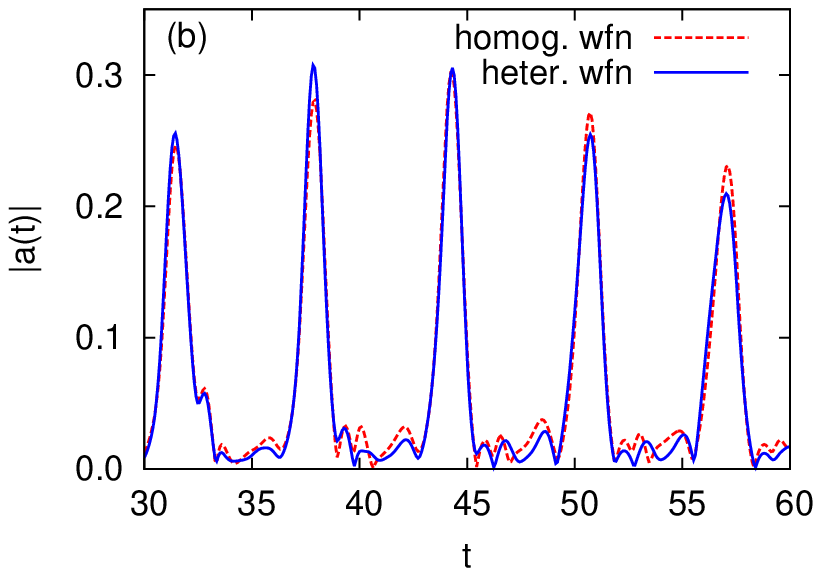}
        \includegraphics[width=8.5cm]{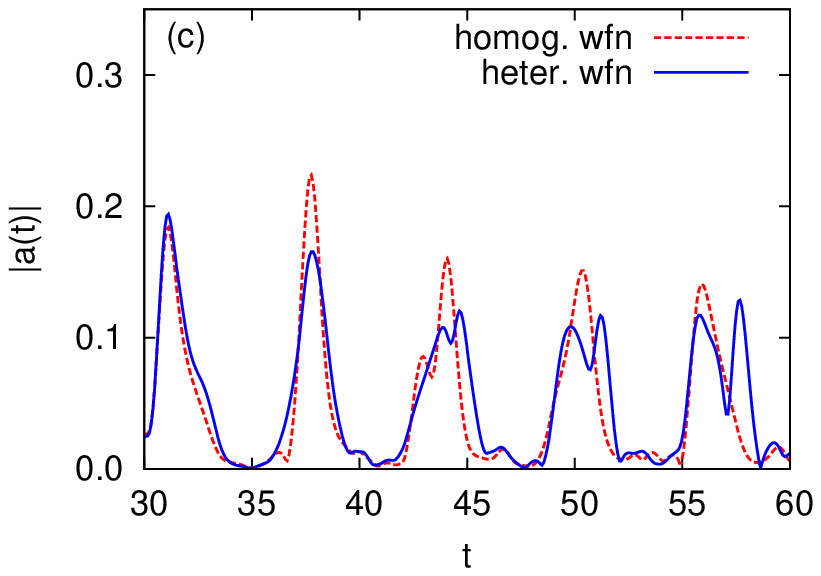}
        \includegraphics[width=8.5cm]{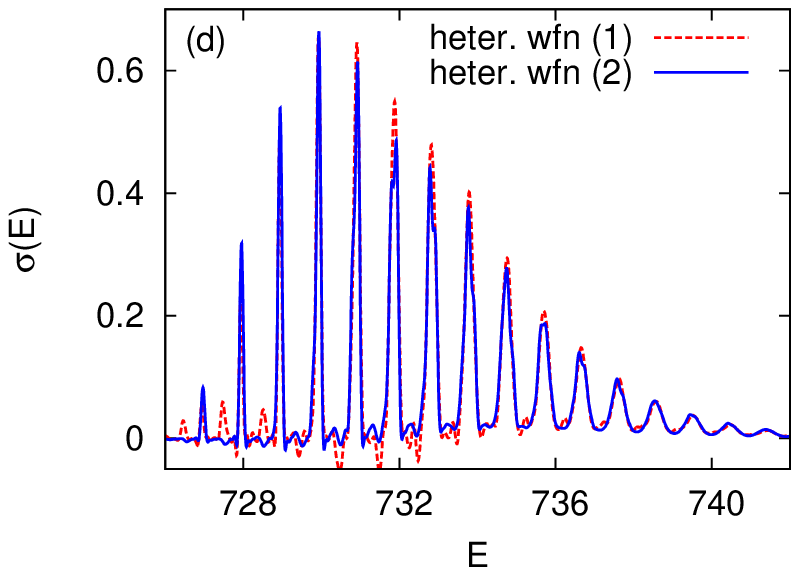}
      \end{center}
     \caption{\figfoot}
     \label{fig:huge}
    \end{figure}

  \placefig{fig:huge}

  Fig.~\ref{fig:huge}a shows the autocorrelation function of the two decoupled
  reference calculations. In both cases the wavefunction consists of 5 SPFs for
  each logical coordinate at all layers, which results in $2\,326\,520$ TD
  coefficients. Since there is no coupling term in the Hamiltonian between the two
  initially displaced DOF, their distance along the tree structure should be
  irrelevant, and in fact both simulations yield nearly identical results.
  When the interaction is turned on, however, the distance between two coupled
  DOF along the tree does matter. In case (2) the tree distance between DOF
  $q_{729}$ and $q_{730}$ is the shortest possible.  Fig.~\ref{fig:huge}b shows
  the autocorrelation of two propagations for case (2). One of them is based on
  the same wavefunction as the reference results.  A second propagation is
  based on a wavefunction with 20 SPFs for the combined mode containing
  $q_{729}$ and $q_{730}$, $Q^{6;2,2,2,2,2}_2$, and 15 SPFs for the two
  neighboring combined modes $Q^{6;2,2,2,2,2}_1$ and $Q^{6;2,2,2,2,2}_3$. One
  layer above the wavefunction has 10 SPFs for logical coordinates
  $Q^{5;2,2,2,2}_2$ and its two neighbors $Q^{5;2,2,2,2}_1$ and
  $Q^{5;2,2,2,2}_3$. Further up in the tree, logical coordinates containing
  $q_{729}$ and $q_{730}$ and its two neighbor coordinates have 7 SPFs each.
  All other logical coordinates at all layers are represented by 4 SPFs. This
  scheme results in $1\,853\,310$ TD coefficients. As seen in
  Fig.~\ref{fig:huge}b, both wavefunctions yield very similar autocorrelations
  until the end of the simulation.  By inspecting the natural populations at
  the final time for both case (2) simulations one notices however that the
  convergence at the lowest layer is rather poor for the homogeneous
  wavefunction with 5 SPFs overall, with a population of the order of $10^{-2}$
  for the last natural orbital of coordinate $Q^{6;2,2,2,2,2}_2$.  When
  inspecting the populations of the propagation with more SPFs for logical
  coordinates close to the displaced mode $Q^{6;2,2,2,2,2}_2$, substantially
  improved values of at least $10^{-4}$ are obtained for all logical
  coordinates containing $Q^{6;2,2,2,2,2}_2$ at all levels.

  Example (1), in which the distance between both initially displaced DOF
  $q_{486}$ and $q_{487}$ is the largest possible within the tree, has a
  different convergence behavior. Fig.~\ref{fig:huge}c compares
  two autocorrelations for case (1). A first propagation is based in
  the same homogeneous wavefunction already discussed for case
  (2). A second simulation is based on a wavefunction with 15 SPFs for
  logical coordinates 
  $Q^{6;1,3,3,3,3}_3$ and $Q^{6;2,1,1,1,1}_1$, which respectively contain DOF
  $q_{486}$ and $q_{487}$, and with 7 SPFs for logical coordinates containing
  also $q_{486}$ and $q_{487}$ at all layers.  All other logical coordinates
  have 4 SPFs and total number of TD coefficients is now $1\,810\,940$. Now,
  both autocorrelations start to diverge after about 30 time units, and
  they also start to diverge from the autocorrelations of case (2) after about
  the same propagation time.  Interestingly however, the least populated
  natural orbitals for logical coordinates $Q^{6;1,3,3,3,3}_3$ and
  $Q^{6;2,1,1,1,1}_1$ have populations of the order of $10^{-3}$ to $10^{-4}$,
  and this is substantially improved to $10^{-4}$ to $10^{-5}$ for the
  non-homogeneous wavefunction. All other natural populations in the whole tree
  structure, for both wavefunctions, remain in the order of $10^{-4}$ or below.
  The spectra of cases (1) and (2) for the non homogeneous wavefunctions
  (Fig.~\ref{fig:huge}d) agree to each other rather well. The differences in
  the autocorrelations are however reflected in a appearance of a few spurious
  lines with low intensity in the low energy part of the spectrum for case (1).

  The analysis of case (2) indicates that when correlated DOF are close to
  each other in the tree structure, i.e. they are connected in the deeper
  layers of the tree, it is easy to achieve convergence by increasing the
  number of SPFs close to such DOF in the tree structure. Even most
  importantly, one can expect an early convergence to the correct trend, since
  even with not very good natural populations at low layers the results are
  already correct. Case (1), on the contrary, is an example of the wrong choice
  of tree structure.  Even with natural populations smaller than case (2) for
  the homogeneous wavefunction, and populations of the order of $10^{-4}$ in
  most layers, the results are different from case (2) and also differ between
  the two different wavefunctions tried for case (1).  In case (1) a general
  and substantial increase of SPFs for the whole tree structure would probably
  be needed to obtain better converged results. The early convergence of the method
  with the number of SPFs appears to be damaged in case (1) in light of the
  difference in results of for the two different wavefunctions tried.

  \subsection{Pyrazine} \label{sec:pyr}

  We now turn our attention to the photophysics of pyrazine using the 24D
  second-order vibronic-coupling Hamiltonian of Raab {\em et
  al.}~\cite{raa99:936}. In pyrazine, the spectrum obtained from excitation to
  the second excited electronic state $S_2$ presents a broad feature due to the
  fast decay of the system into the $S_1$ electronic state through a conical
  intersection. Such a decay occurs during a few ten fs after
  photoexcitation. Although obtaining the right position and approximate shape
  of the broad band is relatively straightforward and several different
  approaches can claim to have achieved it, reproducing the fine details of the
  absorption spectrum is quite hard, since a good quality propagation up to
  relatively long times is needed.  Therefore, the 24D model of pyrazine in
  Ref.~\cite{raa99:936} has often been used to benchmark different quantum
  dynamical approaches including semiclassical methods~\cite{tho00:10282},
  the multiple-spawning method~\cite{ben02:439}, the coupled coherent state
  method~\cite{sha04:3563,sha10:244111}, the matching-pursuit/split-operator
  Fourier-transform method~\cite{che06:124313}, and the
  Gaussian-MCTDH method~\cite{bur08:174104}.

\begin{figure}[h!]
      \begin{center}
        \includegraphics[width=8.5cm]{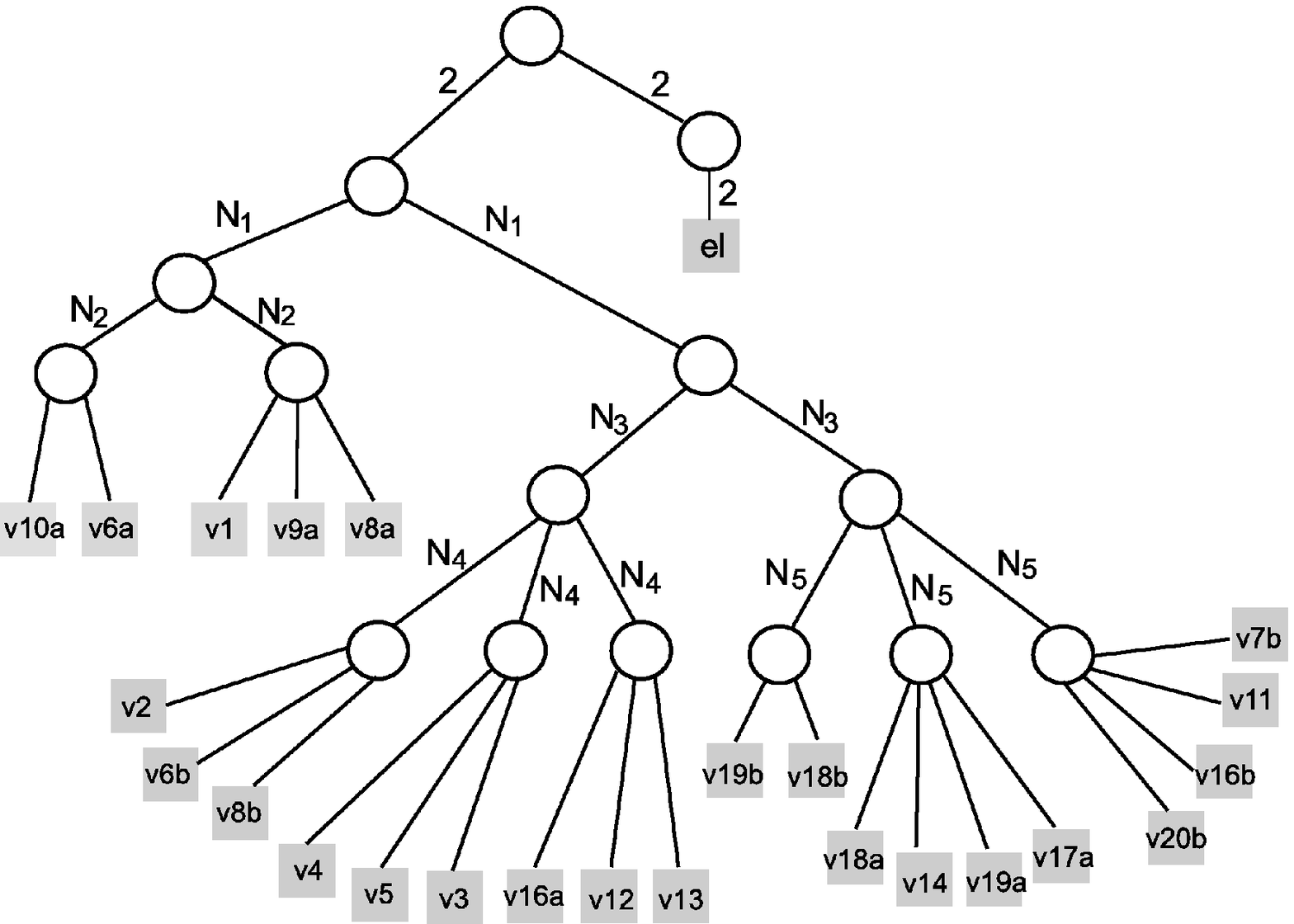}
      \end{center}
     \caption{\figfoot}
     \label{fig:treepyr}
    \end{figure}

  In the 24D simulations reported here, we use the same DVR and grid points as
  in the MCTDH simulations of Refs.~\cite{raa99:936} and \cite{bur08:174104}.
  These MCTDH calculations were based on mode combination scheme in which the
  24 primitive coordinates were grouped into the 8 combined modes
  $Q_1=[\nu_{10a},\nu_{6a}]$,
  $Q_2=[\nu_{1},\nu_{9a},\nu_{8a}]$,
  $Q_3=[\nu_{2},\nu_{6b},\nu_{8b}]$,
  $Q_4=[\nu_{4},\nu_{5},\nu_{3}]$,
  $Q_5=[\nu_{16a},\nu_{12},\nu_{13}]$,
  $Q_6=[\nu_{19b},\nu_{18b}]$,
  $Q_7=[\nu_{18a},\nu_{14},\nu_{19a},\nu_{17a}]$ and
  $Q_8=[\nu_{20b},\nu_{16b},\nu_{11},\nu_{7b}]$.
  A description of these vibrational modes is found in
  Ref.~\cite{raa99:936} and it is beyond the purpose of this paper to reproduce
  their discussion here.  All but one of the reported ML-MCTDH simulations are
  based on this set of combined modes, which are then grouped in the ML tree as
  depicted in Fig.~\ref{fig:treepyr}.  This makes the ML-MCTDH calculations
  numerically easier to compare to previous MCTDH calculations, for example
  when computing wavefunction overlaps, since the ML-MCTDH wavefunction can be
  easily expanded into the corresponding MCTDH form (if the expanded
  form is still of a manageable size) in a similar way as an MCTDH wavefunction
  can be expanded to a standard wavefunction on the primitive grid,
  Eq.~(\ref{eq:ansatz1}).  In the
  simulation named ML-MCTDH-1, however, we further divide modes $Q_7$ and
  $Q_8$, which contain coordinates with large primitive grids, in order to
  better exploit the capabilities of ML-MCTDH. We used  $Q_1 \cdots Q_5$
  as above but:
  $Q_6=[\nu_{19b}]$, $Q_7=[\nu_{18b}]$,
  $Q_8=[\nu_{18a},\nu_{14}]$,  $Q_9=[\nu_{19a},\nu_{17a}]$,
  $Q_{10}=[\nu_{16b}]$, and $Q_{11}=[\nu_{20b},\nu_{11},\nu_{7b}]$.
  This division is accomplished by adding a further (a sixth) layer to the
  tree depicted in Fig.~\ref{fig:treepyr}.

  \placefig{fig:treepyr}
  For non-adiabatic systems with more than one electronic state MCTDH is often
  used in its multi-set formulation~\cite{bec00:1,mey03:251,mey09:book},
  in which different sets of SPFs are used for each electronic DOF. All
  pyrazine MCTDH calculations discussed here were done using the multi-set
  formalism. In the multi-set formalism the SPFs of each state are optimal for
  that state, so that a smaller number of SPFs per state are needed than in a
  single-set calculation.  The computational cost grows exponentially with the
  product of SPFs in each state, but linearly with the number of SPFs in
  different electronic states, which makes multi-set calculations often
  advantageous with respect to single-set ones for the same system.
  In ML-MCTDH, a multi-set formulation would be extremely cumbersome because
  one would end up with several tree structures to be specified, one for each
  electronic state, and a much more complex algorithm. On the other hand, the
  favorable scaling of ML-MCTDH with the number of DOF makes it
  unnecessary to use a multi-set ML formulation.  In simulations of
  non-adiabatic systems with ML-MCTDH the electronic DOF is hence
  just another coordinate that indexes the electronic states, as in usual
  single-set MCTDH calculations, and the ML-MCTDH algorithm remains unchanged.
  Consequently, in constructing the ML tree, one is free to group the
  electronic DOF in any convenient way with other coordinates and set it in any
  level of the tree where it may seem appropriate.  For example, if only a
  subset of coordinates couple the different electronic states and the rest act
  as a bath, it may  make sense to group the electronic DOF with the
  coordinates that control the non-adiabatic coupling, and then the system-bath
  separation can be placed one level above.  In our simulations, the top layer
  contains coordinates ${Q_{vib},q_{el}}$, where $Q_{vib}$ groups the 24
  vibrational modes of the system and $q_{el}$ denotes the electronic DOF.
  $Q_{vib}$ is further divided into two groups of coordinates, one containing
  modes ${\nu_{10a},\nu_{6a},\nu_{1},\nu_{9a},\nu_{8a}}$ and the other one
  containing the rest of modes. Modes ${\nu_{10a},\nu_{6a},\nu_{1},\nu_{9a}}$
  define the "system" in 4D models of pyrazine~\cite{raa99:936}, while the rest
  have been usually termed "bath" modes. Therefore, the first branching of the
  tree after separating electronic and vibrational coordinates can be
  understood as a system-bath separation.  The resulting system and bath
  coordinates are further divided until modes of a manageable size are reached,
  as seen in Fig.~\ref{fig:treepyr}.

  \placetab{tab:pyr}
  Table~\ref{tab:pyr} presents some results for a set of ML-MCTDH
  simulations and three different MCTDH reference calculations. The calculation
  named here MCTDH-1 was reported in Ref.~\cite{raa99:936}. The MCTDH-2 result
  was an MCTDH reference calculation in Ref.~\cite{bur08:174104}, while the
  MCTDH-3 calculation is a new reference result for this system and corresponds
  to the largest MCTDH calculation reported for pyrazine to date.
  All simulations were run up to a propagation time of 150 fs.
  %

\begin{figure}[h!]
      \begin{center}
        \includegraphics[width=8.5cm]{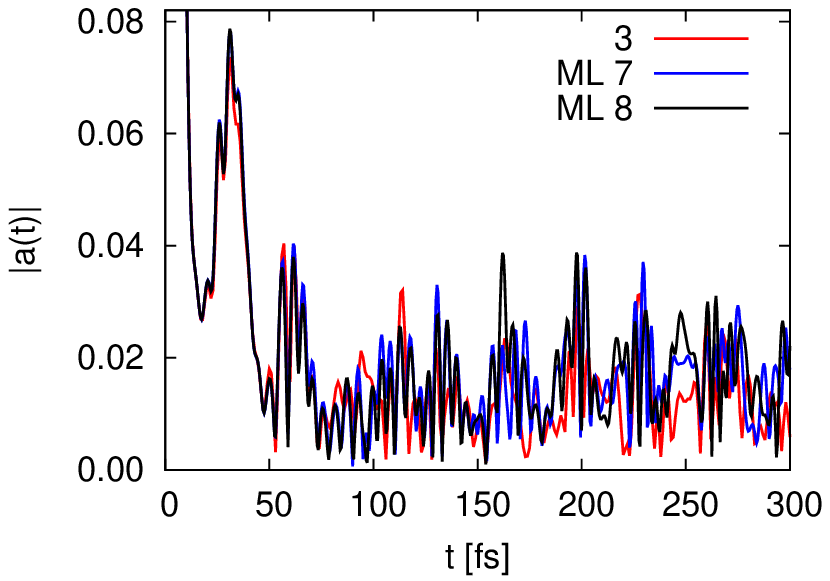}
        \includegraphics[width=8.5cm]{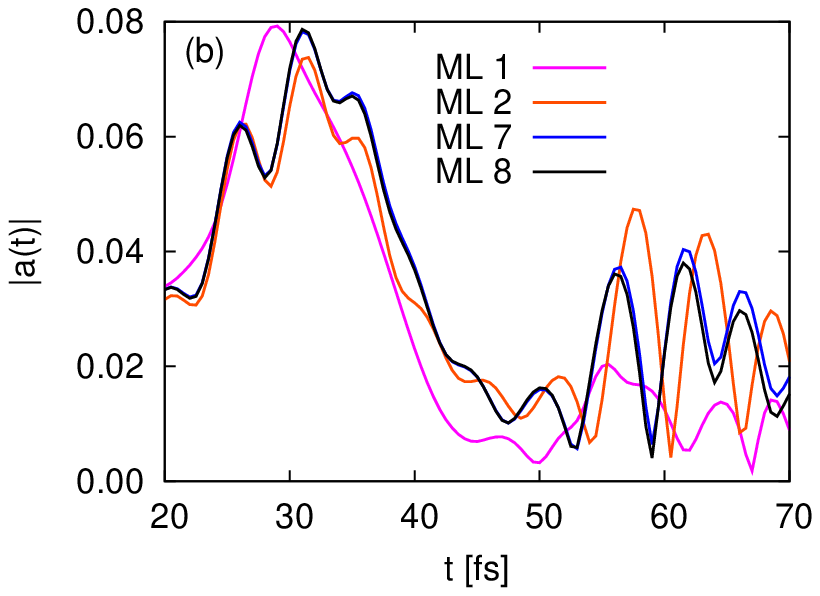}
        \includegraphics[width=8.5cm]{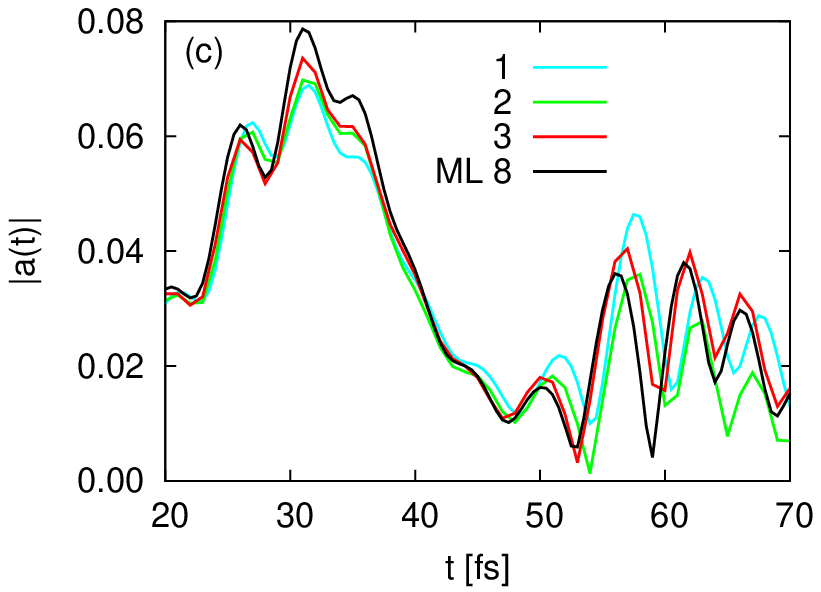}
      \end{center}
     \caption{\figfoot}
     \label{fig:pyrauto}
    \end{figure}

\begin{figure}[h!]
      \begin{center}
           \includegraphics[width=8.5cm]{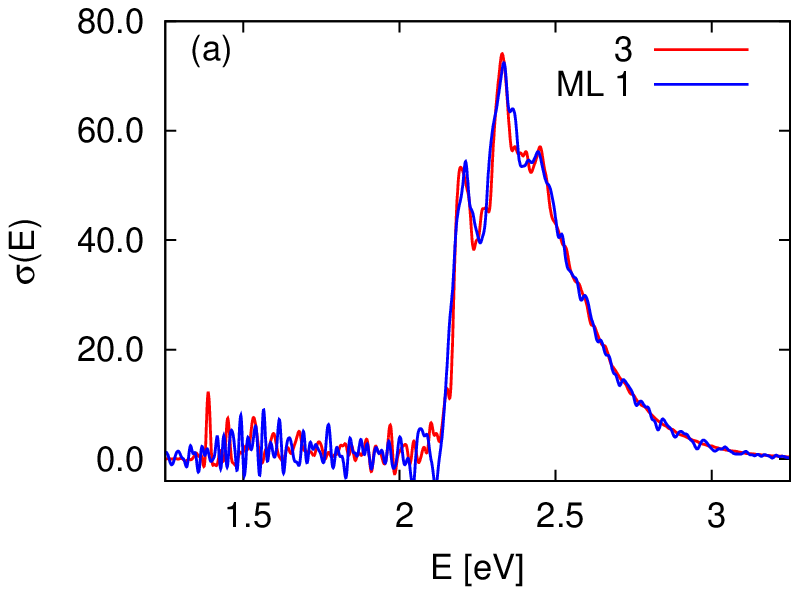}
           \includegraphics[width=8.5cm]{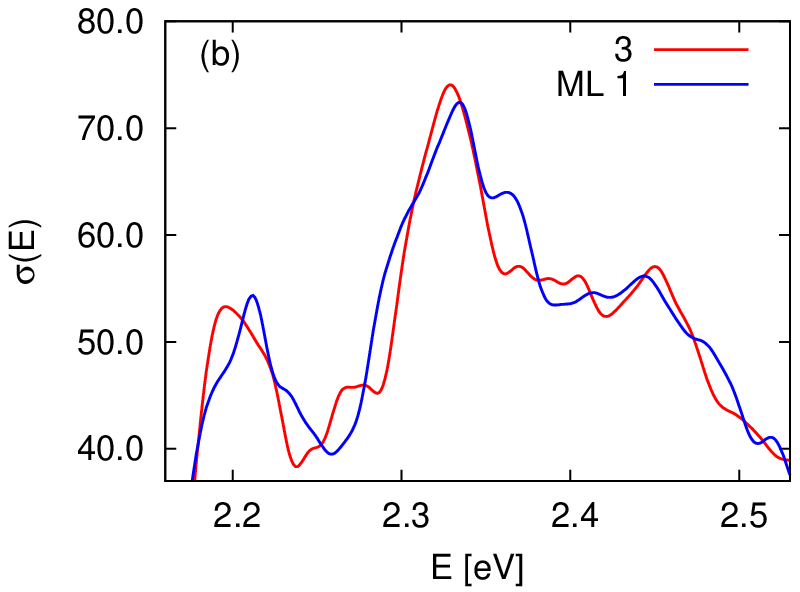}
      \end{center}
     \caption{\figfoot}
     \label{fig:pyrspec_fast}
    \end{figure}

\begin{figure}[h!]
      \begin{center}
           \includegraphics[width=8.5cm]{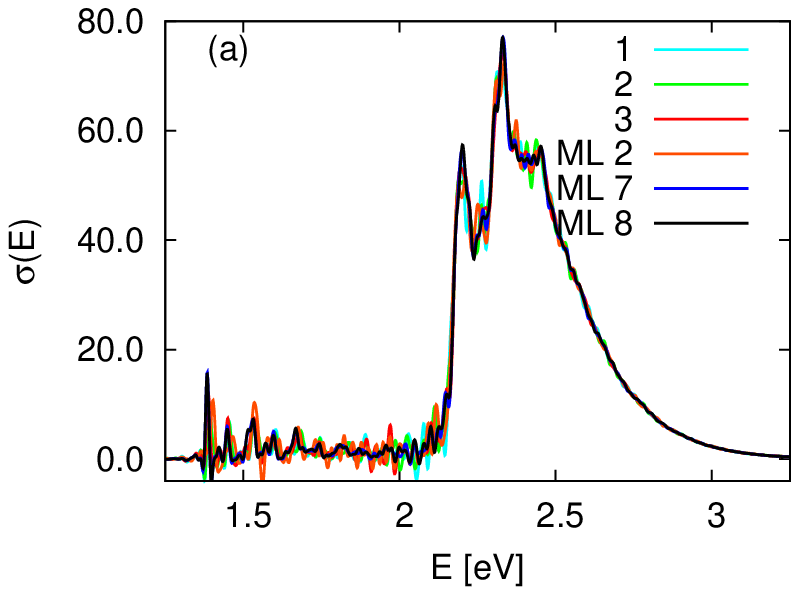}
           \includegraphics[width=8.5cm]{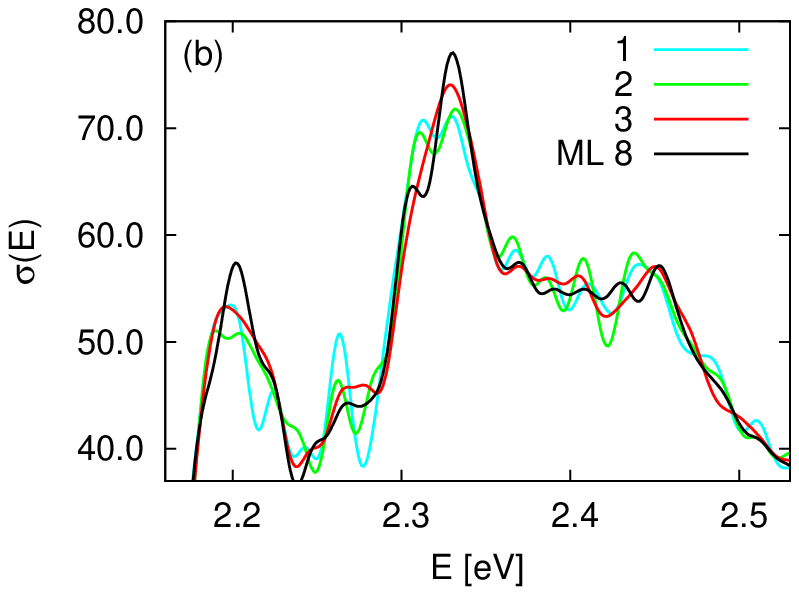}
           \includegraphics[width=8.5cm]{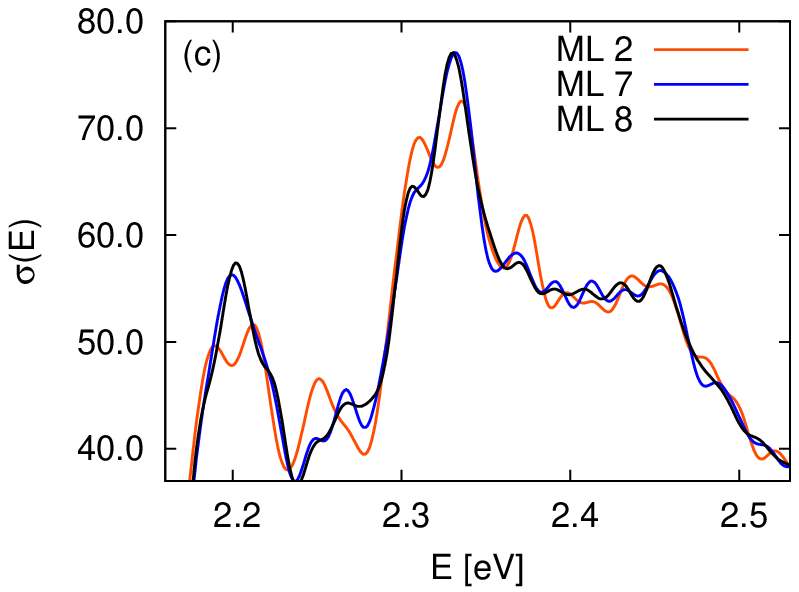}
      \end{center}
     \caption{\figfoot}
     \label{fig:pyrspec}
    \end{figure}

  % Quality & comparison of calculations
  \placefig{fig:pyrauto}
  \placefig{fig:pyrspec_fast}
  Fig.~\ref{fig:pyrauto} presents the absolute value of the autocorrelation
  function for the ML-MCTDH and MCTDH calculations in Table~\ref{tab:pyr}. As
  in the HH cases above, the autocorrelation is computed using the
  relation for real initial states and symmetric Hamiltonians
  $a(t)=\langle\Psi^*(t/2)|\Psi(t/2)\rangle$.
  In Fig.~\ref{fig:pyrauto}b one sees that the ML-MCTDH-1 simulation presents
  severe deficiencies in $|a(t)|$, only following the general shape of the
  initial decay and the small revivals between 25 and 45 fs. This simulation is
  however extremely fast, needing only 7 minutes to run and consuming an
  incredibly small amount of resources. Regarding peak positions and
  intensities, the general features of the spectrum in Fig.~\ref{fig:pyrspec_fast}a,
  although not too accurate, are there. The details of the peaks below
  2.1 eV are not reproduced and the high energy tail, above 2.5 eV, shows
  an artificial oscillatory behavior. But beside this, the main peak is
  amazingly well reproduced, considering the inexpensiveness of the
  calculation. The larger ML-MCTDH-2 simulation, needing
  33 hours (five times faster than the MCTDH-1 reference result), already
  presents the correct trends in the autocorrelation function up to 70 fs,
  resulting in a much improved spectrum (Figs.~\ref{fig:pyrspec}a and
  \ref{fig:pyrspec}c). The autocorrelations of the two most accurate ML-MCTDH
  simulations present features that resemble closely those of the most accurate
  MCTDH result. In fact, when comparing the various structures of $|a(t)|$ in
  Fig.~\ref{fig:pyrauto}c, the tendency of the MCTDH-1, -2, and -3 simulations
  is towards the ML-MCTDH-8 simulation, which seems to indicate that the best
  ML-MCTDH results are better converged than the best MCTDH run. This can be
  seen in the height of the structure between 30 and 40 fs, and also in the
  position of the three peak structures between 55 and 70 fs.
  When turning to the spectra in Fig.~\ref{fig:pyrspec}, similar conclusions
  can be drawn, for example examining closely the peaks at 2.2 and 2.35 eV in
  Fig.~\ref{fig:pyrspec}b. It is remarkable that simulation ML-MCTDH-7, which
  seems to be already better converged than the best MCTDH result reported,
  required about 25\% of runtime and 1\% of memory (in terms of wavefunction
  size) than the latter.

  \placefig{fig:pyrspec}
  In terms of efficiency, the second column in Table~\ref{tab:pyr} contains
  the wall-clock time of each calculation. All ML-MCTDH calculations were
  run on a single-processor, so CPU time equals in this case wall-clock time,
  and used the variable mean-field (VMF)~\cite{bec00:1} propagation scheme.
  The MCTDH calculations were run using the shared-memory parallelized MCTDH
  code~\cite{bri08:141,bri09:67} and 8 processors in parallel, using the
  constant mean-field (CMF) propagation
  scheme.~\cite{bec97:113,bec00:1,man06:168} The speed-up factor of the
  MCTDH parallel computations is 2.9, 3.3, and 3.7 for MCTDH-1, -2, and -3,
  respectively, and this factor has been already multiplied to the MCTDH
  wall-clock times, yielding the CPU time
  that would have been taken on a single CPU so that they can
  be readily compared to the ML-MCTDH values.
  These speed-up are reasonable but not particularly good, which is due to
  the relatively small Hamiltonian operator of the pyrazine model and
  better scalings have been obtained, e.g. for the
  Zundel cation H$_5$O$_2^+$~\cite{bri08:141,bri09:67,bri10:147}.
  All calculations were run on the same machine and CPU type
  (see caption of Table~\ref{tab:pyr}).
  All reported simulations, including the HH ones but except for ML-1, were run
  with a rather high integrator precision of $10^{-7}$.  This was done to
  exclude any numerical artifacts and to ensure that all discussed effects
  originate form various sizes of sets of SPFs.  The fast but low-accuracy
  calculation ML-1 was run with an integrator precision of $10^{-5}$. Reducing
  the high integrator precision will speed-up the calculations by factors
  between 1.5 and 2 without introducing visible changes into the spectra.

  The third column in Table~\ref{tab:pyr} contains the total number of
  time-dependent coefficients in the wavefunction representation. Two aspects
  are remarkable here. First, ML-MCTDH wavefunctions are much more compact
  than MCTDH ones. Second, the CPU time divided by the number of coefficients
  is about one order of magnitude larger in ML-MCTDH, this factor remaining
  quite stable along the series of calculations. Therefore, integrating each
  coefficient in ML-MCTDH is harder than in MCTDH due to the natural overhead
  of the ML-MCTDH algorithm.  However, this is compensated by the much smaller
  number of coefficients in the wavefunction. Such scaling effects become more
  and more pronounced with the size of the system, as was seen for the
  HH case, and ML-MCTDH propagations seem to become more efficient
  than MCTDH ones in terms of CPU time for systems larger than about 20 DOFs.
  Simulations ML-MCTDH-2 to -8 could be made even more efficient further
  dividing the largest combined modes at the lowest layer in a similar
  way as it was done for ML-MCTDH-1.
  Running on 8 processors, MCTDH-3 needed about 1000 hours to complete
  (unscaled wall-clock time), keeping up with the wall-clock time of the best
  ML-MCTDH calculations through parallelization. The MCTDH-3 calculation
  could in principle be made faster than ML-MCTDH using more hardware.
  However, for a
  molecular system doubling the size of pyrazine, a normal MCTDH calculation
  would be probably not doable, or at least not with a reasonable SPF basis
  size. On the contrary, that case would require an additional layer in a
  ML-MCTDH calculation, and the CPU cost would scale approximately linearly,
  rendering the simulation still doable.
  Regarding the different approaches tried on the pyrazine Hamiltonian over the
  years, ML-MCTDH offers by far the best quality/cost relation.  The ML-MCTDH-2
  simulation, which is already of a reasonable quality, takes about one day on
  a single CPU using little resources.  The various ML-MCTDH simulations on the
  cheaper end yield spectra of a quality that other approaches hardly
  reach~\cite{ben02:439,sha04:3563} or reach only by using a much larger amount
  of CPU and memory resources~\cite{che06:124313,bur08:174104}. And again,
  accepting slightly larger deviations in the spectrum, an ML-MCTDH
  calculation on pyrazine takes only 7 minutes.

\section{Summary and Conclusion}  \label{sec:conclusion}

  We presented the implementation of the ML-MCTDH approach into the Heidelberg
  MCTDH package, which is based on the recursive algorithm proposed by Manthe.
  A discussion of the concept of multilayer was given, and the working equations
  were provided making emphasis on the key points of the approach.
  For a comprehensive
  derivation of the working scheme one should see Ref~\cite{man08:164116}.
  The use of the separable terms in the Hamiltonian in the recursive construction
  of the $h$-matrices and mean-fields at the different layers was discussed here
  with some detail.

  The implementation, numerical performance, and general features of the
  ML-MCTDH approach have been tested by running simulations on the Henon-Heiles
  model and on pyrazine.  For Henon-Heiles, simulations for the 6D, 18D and a
  large 1458D case have been reported.  In the 6D and 18D cases, normal MCTDH
  calculations are performed for identical parameter sets for comparison, and
  for the 6D case MCTDH is more efficient than ML-MCTDH. The situation already
  changes for the 18D simulations, in which ML-MCTDH outperforms MCTDH. With
  respect to the natural populations, which are often used as a
  convergence criterion in MCTDH calculations, it is observed that when
  increasing the number of SPFs at lower layers for a given number of SPFs at
  higher layers, the convergence of the higher layer becomes worse although the
  overall quality may improve. This is due to the fact that the dynamics at the
  higher layers becomes more complex after increasing the size of the basis
  below. Hence, with respect to convergence, all layers have to be monitored
  when changing the basis size in some other layer.

  For pyrazine, we test ML-MCTDH on the second-order vibronic-coupling
  Hamiltonian of Raab {\em et. al}, which has been used in testing several
  quantum dynamical methods. By properly choosing the tree scheme, we report a
  ML-MCTDH simulation that takes {\em only} 7 minutes of CPU time in one
  processor and very few memory, and that recovers the main spectral features
  reasonably well. On the other hand, we report ML-MCTDH simulations which are
  converged to at least the same degree as the best reference MCTDH results
  available, taking only 25\% of the CPU time and 1\% of wavefunction storage
  space with respect to such reference results. As seen in the pyrazine case,
  ML-MCTDH allows for a consistent separation of the degrees of freedom
  considered as system and the ones considered as bath, providing the
  variationally optimal system-bath evolution for the particular selection of
  the basis size and the layering scheme. By using ML-MCTDH for system-bath
  problems, schemes that simplify the system-bath {\em
  ansatz}~\cite{lop10:104103} or schemes that reduce a larger bath to a subset
  of effective modes~\cite{gin06:144103}, become unnecessary.

  ML-MCTDH is an efficient tool for model systems, which had already been
  demonstrated by works of Wang and Thoss. ML-MCTDH has been shown here to be a
  very powerful tool to treat more realistic molecular Hamiltonians as well.
  The fact that a tree structure needs to be chosen, often based on intuition
  or experience, and that convergence has to be monitored at the different
  layers, makes its use less straightforward than MCTDH. Therefore, some work
  in the direction of automating certain decisions regarding the tree structure
  will have to be carried on. Regarding efficiency, it will be interesting to
  develop good dedicated constant mean-field (CMF) propagation schemes
  \cite{bec97:113,bec00:1,man06:168} for the ML case.  And, of course,
  parallelizing the ML-code will be an important step, as ML-MCTDH is for
  treating large systems.
  Beyond applications of ML-MCTDH to problems involving nuclear dynamics, for
  which it has certainly a large potential, new directions will have to be
  explored in the future.  We believe that methods based on or taking advantage
  of the ML-MCTDH concept can be useful in a wide range of situations. For
  example, we envisage that in simulations of mixtures of different kinds of
  particles (different kinds of fermions and bosons, electrons and nuclei in
  molecules, etc.), the different types of particles can be separated and
  correlated to each other at the top layer, while the internal dynamics of
  each group takes place in the layers below.
  ML-MCTDH constitutes a consistent and powerful tool for the quantum-dynamical
  description of high dimensional molecular systems. This and other kinds of
  challenging applications remain the subject of future investigations.

\section{Acknowledgments}

    Financial support by the Deutsche Forschungsgemeinschaft (DFG) is
    gratefully acknowledged.

%\bibliography{refs}
%\bibliographystyle{aip}

\clearpage  \pagestyle{plain}
\clearpage  \pagestyle{empty}
        \begin{table}[t]
         \caption{
         Simulation parameters of the various MCTDH and ML-MCTDH 24D
         pyrazine calculations.  The second column contains the wall-clock time
         of each simulation. ML-MCTDH calculations were run on a single CPU, so
         that the wall-clock time equals the CPU time.  The MCTDH calculations
         were run on 8 CPUs using shared-memory
         parallelization~\cite{bri08:141,bri09:67}. The speed-up factor of the
         8-processor parallel pyrazine calculations is 2.9, 3.3 and 3.7 for the
         MCTDH-1, -2 and -3 cases, respectively. The wall-clock times given for
         the MCTDH reference results are already scaled up by the corresponding
         speed-up factor and therefore reflect the time that such simulation
         would have taken on a single processor. Therefore they can be readily
         compared to the ML-MCTDH values.  All simulations were run on the same
         machine and CPU type, namely Quad-Core AMD Opterons, processor type
         2384 running at 2.7 GHz.  The third column shows the total number of
         time-dependent coefficients propagated in each case. The fourth column
         contains, for ML-MCTDH simulations, the number of SPFs for each node
         of the tree according to the representation in Fig.~\ref{fig:treepyr}.
         The parenthesis indicate that different $N_5$ values were taken for
         each of the branches.  The asterisk for the ML-1 case indicates that
         there is a further layer below the branches corresponding to $N_5$
         (See text and Fig.~\ref{fig:treepyr} for details).  For the MCTDH
         calculations, the fourth column contains in each parenthesis the
         number of SPFs for combined modes $Q_1$ to $Q_8$ for electronic states
         $S_1$ and $S_2$, respectively.
         All reported simulations, except for ML-1, were run with a rather high
         integrator precision of $10^{-7}$. (See text). ML-1 was run with
         $10^{-5}$. Reducing the high integrator precision will speed-up the
         calculations by factors between 1.5 and 2 without introducing visiable
         changes into the spectra.}
         \label{tab:pyr}
        \end{table}

\clearpage
        \makebox[\textwidth]{%
         \begin{tabular}{l@{\hspace{0.5cm}}
                         r@{\hspace{0.5cm}}
                         rc}
            \hline
            \hline
         Simulation     &  CPU time [h:m]  &     tot. coef   & [$N_1$,$N_2$,$N_3$,$N_4$,$N_5$] \\
            \hline
            ML-1        &  0:07           &      $22\,444$  & [$4,4,3,2,(3,2,2)^*$]    \\
            ML-2        &  33:48          &     $206\,660$  & [$12,13,10,8,8$] \\
            ML-3        &  57:12          &     $238\,054$  & [$14,14,11,8,(12,8,8)$] \\
            ML-4        &  113:54         &     $294\,820$  & [$16,16,13,10,(13,10,10)$] \\
            ML-5        &  135:31         &     $288\,324$  & [$25,25,11,8,(12,8,8)$] \\
            ML-6        &  252:31         &     $337\,408$  & [$25,25,13,10,(13,10,10)$] \\
            ML-7        &  725:29         &     $456\,584$  & [$30,30,18,12,(15,12,13)$] \\
            ML-8        &  1123:38        &     $512\,560$  & [$32,32,21,12,(17,12,14)$] \\
            \hline
                        &                 &                 & [$(n_1^{S_1},n_1^{S_2}),\ldots,(n_8^{S_1},n_8^{S_2})$]   \\
            \hline
            MCTDH-1     &   155:58        &  $3\,029\,424$  & [$(14,11),(8,7),(6,5),(6,4),(4,5),(7,7),(5,5),(3,4)$] \\
            MCTDH-2     &   629:58        & $11\,282\,152$  & [$(19,12),(10,7),(5,4),(7,4),(5,3),(11,9),(7,5),(4,4)$] \\
            MCTDH-3     &  3721:40        & $46\,351\,392$  & [$(21,15),(12,8),(7,6),(8,5),(7,5),(12,10),(7,5),(5,5)$] \\
            \hline
            \hline
         \end{tabular}
        }

\clearpage
 \section*{Figure Captions}
  \pagestyle{plain}
  \figcaption{fig:tree6}{
    Tree structures for the MCTDH and ML-MCTDH wavefunctions of the 6D
    Henon-Heiles simulations.  (a) MCTDH wavefunction tree, in which the
    coordinates are combined in groups of two. $N_2$ refers in this particular
    case to the number of primitive basis functions or grid points.
    (b) ML-MCTDH wavefunction tree.
    This tree is similar to the MCTDH tree, but the three combined modes have
    been separated by adding an extra layer. For the ML-MCTDH wavefunction
    $N_3$ corresponds now the number of primitive basis functions and for
    $N_2\equiv N_3$ (and same $N_1$) the ML-MCTDH case becomes numerically
    identical to the MCTDH one.}

\begin{figure}[h!]
      \begin{center}
        \includegraphics[width=8.5cm]{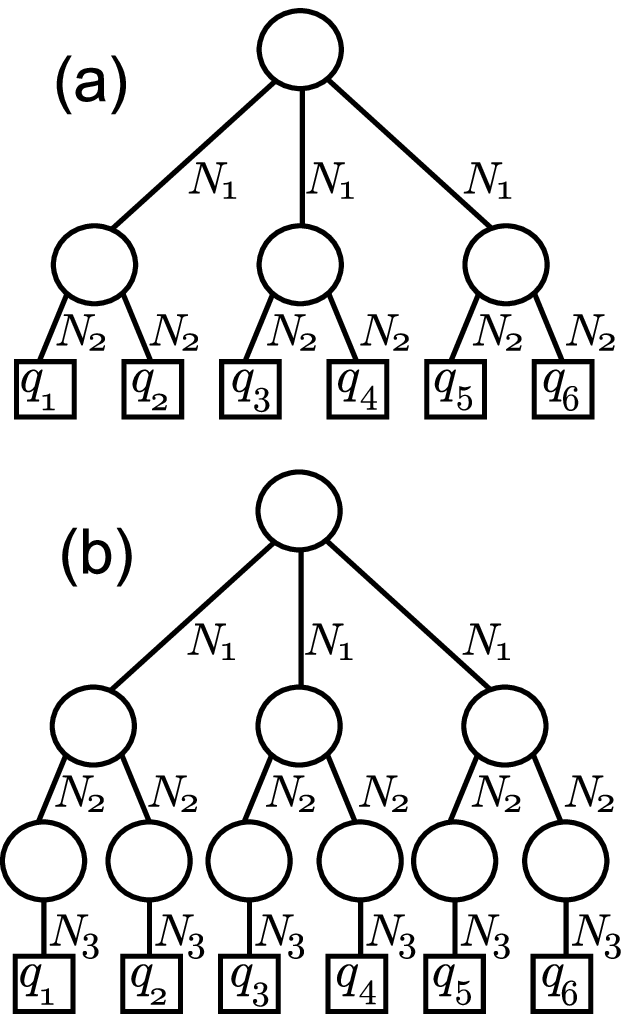}
      \end{center}
     \caption{\figfoot}
     \label{fig:tree6}
    \end{figure}

  \figcaption{fig:auto6D1}{
     (color) (a) Absolute value of the autocorrelation function for various
     MCTDH and ML-MCTDH 6D Henon-Heiles simulations with a coupling parameter
     $\lambda = \lambda_0$. The numbers in parenthesis correspond to
     ($N_1,N_2$) in Fig.~\ref{fig:tree6}. In the key, plain numbers indicate
     MCTDH calculations and the symbol ML designates ML-MCTDH calculations.
     (b) Difference between the autocorrelations of the ML-MCTDH simulations
     and the (50,24) MCTDH calculation.}

  \figcaption{fig:auto6D2}{
     (color) (a) Absolute value of the autocorrelation function for various
     MCTDH and ML-MCTDH 6D Henon-Heiles simulations with a coupling parameter
     $\lambda = 2\lambda_0$.  The numbers in parenthesis correspond to
     ($N_1,N_2$) in Fig.~\ref{fig:tree6}.  (b) Difference between the
     autocorrelations of the ML-MCTDH simulations and the (50,24) MCTDH
     calculation. In the key, ML is for ML-MCTDH.}

  \figcaption{fig:tree18}{
     Tree structures for the MCTDH and ML-MCTDH wavefunctions of the 18D
     Henon-Heiles simulations. The MCTDH wavefunction (a) consists of six
     combined modes, each of them grouping three primitive coordinates. The
     ML-MCTDH wavefunction (b) starts dividing the system in three logical
     coordinates, and each of them is further divided in three combined modes.
     The resulting tree has three layers of TD coefficients.}

\begin{figure}[h!]
      \begin{center}
        \includegraphics[width=8.5cm]{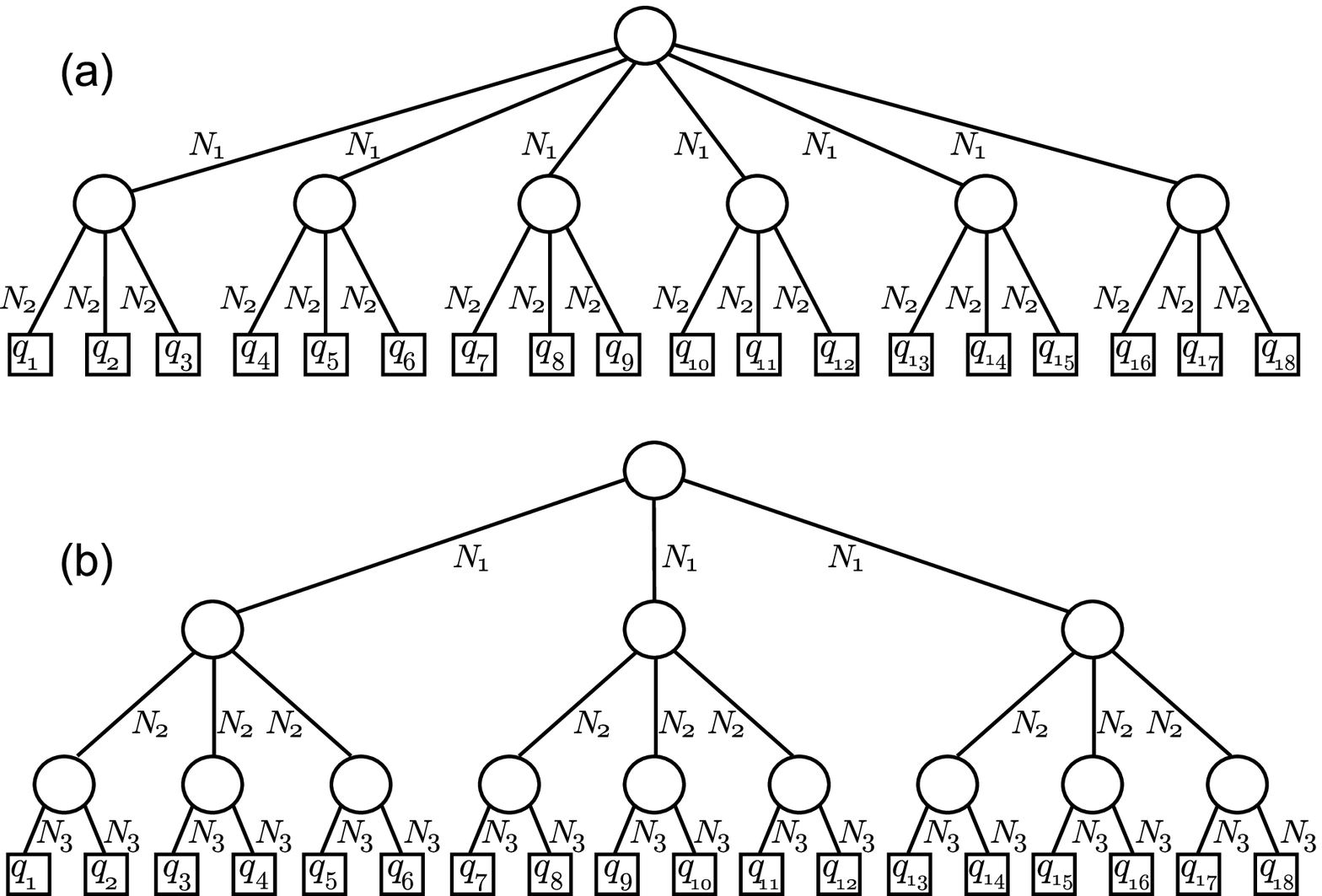}
      \end{center}
     \caption{\figfoot}
     \label{fig:tree18}
    \end{figure}

  \figcaption{fig:auto18D}{
     (color) (a) Absolute value of the autocorrelation function for various
     MCTDH and ML-MCTDH 18D Henon-Heiles simulations with a coupling parameter
     $\lambda = \lambda_0$. For the MCTDH cases the number in parenthesis
     refers to $N_1$ in Fig.~\ref{fig:tree18}a. For the ML-MCTDH cases the
     number in parenthesis refers to $N_1$ and $N_2$ in Fig.~\ref{fig:tree18}b,
     which are set equal in the reported simulations.  (b) Detailed view of the
     last recurrence structure between 55 and 60 time units. In the key, ML is
     for ML-MCTDH.}

  \figcaption{fig:auto18DSI}{
     (color) (a) Absolute value of the autocorrelation function for various
     MCTDH and ML-MCTDH 18D Henon-Heiles simulations with a coupling parameter
     $\lambda = 2\lambda_0$. For the MCTDH cases the number in parenthesis
     refers to $N_1$ in Fig.~\ref{fig:tree18}a. For the ML-MCTDH cases the
     number in parenthesis refers to $N_1$ and $N_2$ in Fig.~\ref{fig:tree18}b,
     which are equal in the reported simulations. (b) Detailed view of the
     recurrence structures between 10 and 30 time units.  In the key, ML is for
     ML-MCTDH.}

\begin{figure}[h!]
      \begin{center}
        \includegraphics[width=8.5cm]{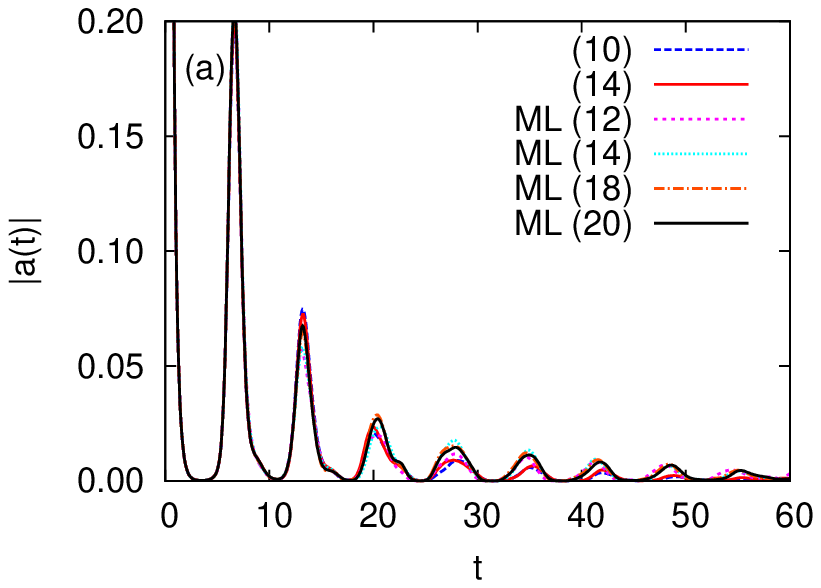}
        \includegraphics[width=8.5cm]{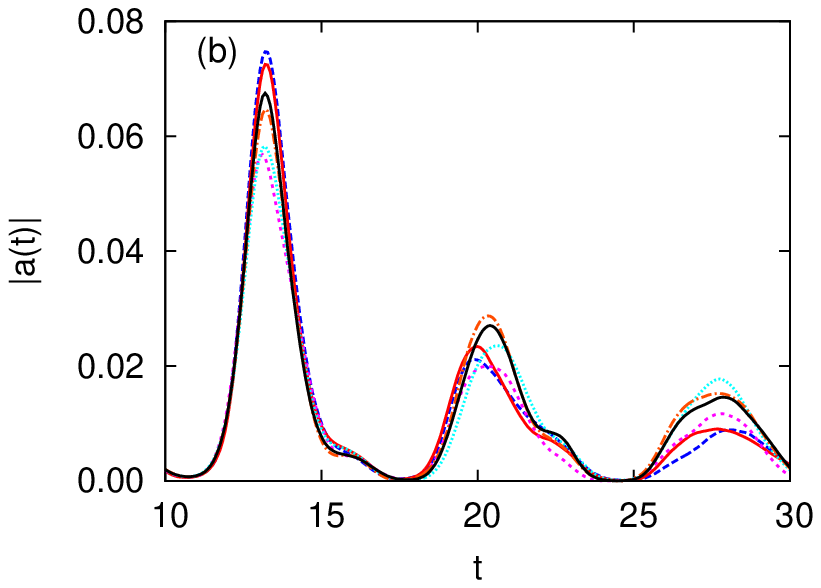}
      \end{center}
     \caption{\figfoot}
     \label{fig:auto18DSI}
    \end{figure}

  \figcaption{fig:spec18DSI}{
     (color) (a) Spectra for various MCTDH and ML-MCTDH 18D Henon-Heiles
     simulations with a coupling parameter of $\lambda = 2\lambda_0$. The
     numbers in parenthesis have the same meanings as in
     Fig.~\ref{fig:auto18DSI}. (b) Detailed view of the three first peaks
     between $6.2$ and $7.9$ energy units. In the key, ML is for ML-MCTDH.
     The spectra are obtained by a Fourier-transform of the autocorrelation
     function using a $cos^2$ filter~\cite{bec00:1,mey09:book} to minimize
     spurious effects known as Gibbs phenomenon.}

  \figcaption{fig:huge}{
     (color) Absolute value of the autocorrelation of the ML-MCTDH 1458D
     Henon-Heiles simulations for time units 30 to 60.  (a) Reference results
     without coupling term between the two displaced coordinates for case (1)
     (red dashed), in which the two initially displaced degrees of freedom
     belong to different logical coordinates at all layers, and case (2) (blue
     line), in which the two initially displaced degrees of freedom belong to
     the same logical coordinate at all layers.
     (b) For case (2), homogeneous wavefunction (red dashed) and non
     homogeneous wavefunction with more SPFs at low layers (blue line).
     (c) For case (1), homogeneous wavefunction (red dashed) and non
     homogeneous wavefunction (blue line).
     The pairs of autocorrelation functions presented in plots (b) and (c) are
     almost identical until about 15 time units, but the autocorrelation
     functions of case (1) and (2) start to differ already after 5 time units.
     These differences increase with time and can be inspected for the time
     interval 30 to 60 by comparing figure (b) with (c).
     (d) Spectra of the calculations with the non homogeneous wavefunctions
     for cases (1) and (2).  The spectra are obtained by a Fourier-transform
     of the autocorrelation function using a $cos^2$
     filter~\cite{bec00:1,mey09:book} to minimize spurious effects known as
     Gibbs phenomenon.} 

  \figcaption{fig:treepyr}{
     Tree structure used in most of the ML-MCTDH simulations of 24D pyrazine.
     The maximum depth of the tree is five layers, and the first one separates
     the 24 vibrational coordinates and the discrete electronic degree of
     freedom. The number of SPFs denoted $N_4$ need not be necessarily the
     same, and in fact in some of the ML-MCTDH simulations they are chosen to
     be different. The same is true for the three $N_5$'s, which can be
     different. For the fast ML-1 calculation a 6th layer was added, see text.}

  \figcaption{fig:pyrauto}{
     (color) Absolute value of the autocorrelation function for the 24D
     pyrazine calculations. The numeration of the calculations is consistent
     with Table~\ref{tab:pyr}.  (a) Best MCTDH result and two best ML-MCTDH
     results from 0 to 300 fs.  (b) Detailed view of the recurrences between 20
     and 70 fs for all reported MCTDH simulations and the best ML-MCTDH simulation (b),
     and for the
     two worst and two
     best converged ML-MCTDH runs (c). In the key, ML is for ML-MCTDH.}

  \figcaption{fig:pyrspec_fast}{
    (color) (a) Spectra of the fastest ML-MCTDH 24D pyrazine calculation
    ($2.2\times 10^4$ TD coeff., 7 min. of CPU) and the best reference MCTDH
    result ($4.6\times 10^7$ TD coeff., 2901 hours of CPU). (b) Detailed view
    of the energy domain between 2.1 and 2.6 eV. The spectra are obtained by
    a Fourier-transform of the autocorrelation function using a $cos^2$
    filter~\cite{bec00:1,mey09:book} to minimize spurious effects known as
    Gibbs phenomenon. The numeration of the calculations is
    consistent with Table~\ref{tab:pyr}. In the key, ML is for ML-MCTDH.}

  \figcaption{fig:pyrspec}{
    (color) (a) Spectra of the three reference MCTDH simulations and three of
    the ML-MCTDH simulations, among them the two best results, for 24D
    pyrazine. The spectra are obtained by a Fourier-transform of the
    autocorrelation function using a $cos^2$ filter~\cite{bec00:1,mey09:book}
    to minimize spurious effects known as Gibbs phenomenon.
    The numeration of the calculations is consistent with
    Table~\ref{tab:pyr}.  For the energy domain between 2.1 and 2.6 eV: (b) The
    three reference MCTDH calculations compared to the best ML-MCTDH result and
    (c) ML-MCTDH simulations 2, 7 and 8. In the key, ML is for ML-MCTDH.}

%----------------------------------------------------------------------------
\clearpage

\clearpage

\clearpage

\clearpage

\clearpage

\clearpage

\clearpage

\clearpage

\clearpage

\clearpage

\clearpage

\clearpage

\end{document}